\pdfoutput=1
\documentclass[11pt, reqno,preprint]{article}
\usepackage{jheppub}
\usepackage{epsfig}
\usepackage{amssymb}
\usepackage{amsmath}
\usepackage{bbm}
\usepackage{mathrsfs}
\usepackage{hyperref}
\usepackage{multirow}

\def\be{\begin{equation}}
\def\ee{\end{equation}}
\def\ba{\begin{eqnarray}}
\def\ea{\end{eqnarray}}
\def\nl{\nonumber\\}

\def\Res{{\rm Res\,}}

\def\epsa{\epsilon}  
\def\dDisc{{\rm dDisc}\,}
\def\gpure{g^{\rm pure}}
\def\sreg{{\rm reg}}
\def\slog{{\rm log}}
\def\OO{\mathcal{O}}

\newcommand{\lsim}{\mathrel{\hbox{\rlap{\lower.55ex \hbox{$\sim$}} \kern-.3em \raise.4ex \hbox{$<$}}}}
\newcommand{\gsim}{\mathrel{\hbox{\rlap{\lower.55ex \hbox{$\sim$}} \kern-.3em \raise.4ex \hbox{$>$}}}}

\def\rhobar{\bar{\rho}}
\def\zbar{\bar{z}}
\def\j{J}
\def\c{c}
\def\GG{\mathcal{G}}
\def\comment#1{{[\color{blue}{#1}]}}

\title{Analyticity in Spin in Conformal Theories}

\author{Simon Caron-Huot}
\affiliation[a]{Department of Physics, McGill University, 3600 rue University, Montr\'eal, QC Canada H3A 2T8}
\emailAdd{schuot@physics.mcgill.ca} \abstract{
Conformal theory correlators are characterized
by the spectrum and three-point functions of local operators.
We present a formula which extracts this data as an analytic function of spin.
In analogy with a classic formula due to Froissart and Gribov, it is sensitive only to an ``imaginary part'' which appears after analytic continuation to Lorentzian signature, and it converges thanks to recent bounds on the high-energy Regge limit.
At large spin, substituting in cross-channel data, the formula yields $1/\j$ expansions with controlled errors.
In large-$N$ theories, 
the imaginary part is saturated by single-trace operators. For a sparse spectrum, it manifests the
suppression of bulk higher-derivative interactions that constitutes the signature of a local gravity dual in Anti-de-Sitter space.
}

\begin{document}

\maketitle

\section{Introduction}

The conformal bootstrap has highlighted the power of basic principles
to constrain and even solve quantum field theories.  The program combines inequalities obtained from
conformal symmetry, unitarity and crossing symmetry to quantitatively narrow down the possible
values of observables like scaling exponents and operator product expansion (OPE) coefficients.
The modern numerical bootstrap \cite{Rattazzi:2008pe} has been applied very successfully
to numerous models in various dimensions (far too many to even attempt to review here)
yielding for instance precise critical exponents for
the three-dimensional Ising model \cite{ElShowk:2012ht,El-Showk:2014dwa}.
Of course, the same principles  constrain non-conformal theories as well \cite{Adams:2006sv}, although it remains
challenging at the moment to use them as a quantitative solution method (see \cite{Paulos:2016but} for encouraging steps).

In parallel to numerical advances, an analytic approach has been developed
which implements constraints that are more readily visible in Lorentzian rather then Euclidean signature.
Namely, by focusing on a Lorentzian limit which selects the contribution from operators with large spin,
crossing symmetry predicts weighted averages over their OPE data.
Assuming that individual contributions are sufficiently regular and close to the average,
this then provides an asymptotic $1/\j$ expansion of this data
\cite{Komargodski:2012ek,Fitzpatrick:2012yx,Kaviraj:2015cxa,Kaviraj:2015xsa}.
This assumption has been confirmed in explicit examples, for instance in
the three-dimensional Ising model again, where the resulting expansion appears to remain accurate
all the way down to spin two \cite{Alday:2015ota,Simmons-Duffin:2016wlq}!
Gaining control over the errors in this expansion will be crucial to mesh it with numerical approaches.

Another important application of the analytic bootstrap is to large-$N$ theories with a so-called
sparse spectrum, which are famously conjectured to be dual to weakly coupled theories of gravity in Anti-de-Sitter space
\cite{Heemskerk:2009pn}.  The large-$N$ crossing equation then admit homogeneous solutions
which are in one-to-one correspondence with possible higher-derivative bulk interactions,
and whose coefficients need to be small for the bulk theory to admit a local interpretation.
It has been recognized that this smallness is tied to the good high-energy (Regge) behavior of the theory,
a feature which is particularly apparent in Mellin space \cite{Penedones:2010ue,Fitzpatrick:2011ia,Alday:2016htq,Rastelli:2016nze}.
Physically, a peculiar feature of these higher-derivative solutions is that they have a bounded spin,
which clashes with the experience, reviewed below, that physics should be analytic in spin.

The goal of this paper is to establish the phenomenon of analyticity in spin in conformal field theories,
and to quantify its implications by means of an inversion formula.
In the context of the large spin bootstrap, this formula will explain why the spectrum organizes into analytic families
and provide control over individual OPE coefficients as opposed to averages.
At the same time, by upgrading the asymptotic expansion to a convergent one, it will clarify the sense in which spin two is ``large enough''.  In the context of large-$N$ theories with sparse spectrum, the same formula will
bound the strength of higher-derivative bulk interactions.
In both cases, the validity of the formula will be tied to the good behavior of correlation functions
in the high-energy Regge limit.

\subsection{Why good behavior in the Regge limit is constraining}\label{ssec:toy}

\def\Mi{f}
Physically, analyticity in spin reflects the fact that not any low-energy expansion can resum into
something that is sensible at high energies.

\begin{figure}
\be\begin{array}{cc}
\hspace{0mm}\def\svgwidth{68mm}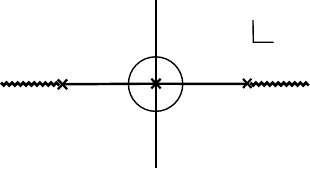
& \hspace{6mm}
\raisebox{17mm}{\resizebox{7mm}{!}{$\Rightarrow$}}
\hspace{6mm}\def\svgwidth{68mm}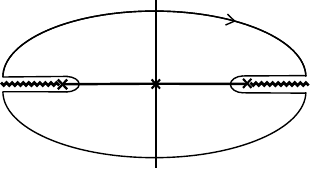
\end{array}\nonumber\ee
\caption{Relation between low-energy coefficients and discontinuity. Analyticity in spin holds if the arcs at infinity
(the Regge limit) can be dropped.}
\label{fig:Eplane}
\end{figure}

Mathematically, this can be illustrated by a  simple single-variable model.
Consider an ``amplitude'' which admits a low-energy Taylor series:
\be
 \Mi(E)= \sum_{\j=0}^\infty \Mi_\j E^\j\,.
 \label{example_f_rho}
\ee
We suppose that we are given the following information: $\Mi(E)$ is analytic except for branch cuts at real energies $|E|>1$,
and $|\Mi(E)/E|$ is bounded at infinity.  (In the physical application below,
$\Mi(E)$ will represent the four-point correlator and its low-energy expansion will be provided by the Euclidean OPE;
at the thresholds $E=\pm 1$ some distances become timelike.)
With the stated assumptions, an elementary contour deformation argument relates the series coefficients
to the discontinuity of the amplitude, as shown in fig.~\ref{fig:Eplane}:
\ba
 \Mi_\j &\equiv& \frac{1}{2\pi i}\oint_{|E|<1} \frac{dE}{E} E^{-\j} \Mi(E)
 \\ &=& \frac{1}{2\pi}\int_1^\infty  \frac{dE}{E} E^{-\j}\,\left({\rm Disc}\,\Mi(E)+(-1)^\j {\rm Disc}\,\Mi(-E)\right) \qquad (\j> 1), \label{example_fj_int}
\ea
where ${\rm Disc}\,f=-i\big[\Mi(E(1+i0))-\Mi(E(1-i0))\big]$. The second line follows from the first
using the assumed high-energy behavior to drop large arcs at infinity.

As a concrete example, one may take the function $\Mi(E)=-\log(1-E^2)$:
upon inserting its discontinuity
${\rm Disc}\,\Mi=2\pi$, the integral indeed produces $\Mi_\j=(1+(-1)^\j)/\j$, as expected.

Now let us focus on a single coefficient, say $\Mi_2$.
It may seem paradoxical that it can be recovered from the discontinuity of $\Mi(E)$,
given that varying $\Mi_2$ alone in eq.~(\ref{example_f_rho}) clearly leaves ${\rm Disc}\,\Mi(E)$ unchanged.
The point is that \emph{given the constraint that $|\Mi(E)/E|$ is bounded at infinity},
the coefficient $\Mi_2$ (or any finite number of coefficients) cannot be varied independently of all the others.
Rather, the coefficients form a much more rigid structure, that is an analytic function of spin.
This is explicited by the integral in eq.~(\ref{example_fj_int}), which defines an analytic function
provided that the real part of $J$ is large enough.
(More precisely, there are two analytic functions, for even and odd spins, reflecting that there are two branch cuts.)

These are the key features of the classic Froissart-Gribov formula \cite{Gribov:1961ex,Collins:1977jy,Donnachie:2002en},
which is conceptually the same but with Legendre functions instead of power laws.
Historically, the Froissart-Gribov formula established the analyticity in spin of partial amplitudes in relativistic $S$-matrix theory,
thus paving the way for phenomelogical applications of Regge theory.

We will show that OPE coefficients in unitary conformal field theories are of a similar type:
they are not independent from each other, but rather organize into rigid analytic functions.
Furthermore, they can be extracted from a ``discontinuity'' which would naively seem to annihilate each individual contribution.

An immediate quantitative implication can be illustrated in the example of large-$N$ theories with a sparse spectrum,
where we will see that the discontinuity is negligible below a gap $\Delta^2_{\rm gap}$.
The preceding formula then gives a result which decays rapidly with spin,
\be
 \Mi_\j \sim  \int_{\Delta_{\rm gap}^2}^\infty \frac{dE}{E}E^{-\j} {\rm Disc}\,\Mi(E) \sim (\Delta_{\rm gap})^{-2\j},
\ee
which in the context of gauge-gravity duality will be interpreted as the expected suppression of higher-derivative corrections,
if the bulk theory is local to distances of order $1/\Delta_{\rm gap}$ times the AdS curvature radius.
Notice the essential role of controlling the Regge limit:
nothing would be learnt from this argument for a given $\j$ if we didn't know that $|\Mi(E)/E^{\j}|$ vanishes at infinity.
Physically, eq.~(\ref{example_fj_int}) and the Froissart-Gribov formula can be regarded as dispersion relations for partial waves,
since their input are discontinuities of amplitudes (this is further discussed in section \ref{ssec:fg0}).

The goal of this paper to obtain similar dispersive representations but which extract
the OPE coefficients in unitary CFTs (projecting out descendants and extracting only primary operators).
Convergence will be established for all spins higher than one, by borrowing
ideas from the recent ``bound on chaos'' as well as from a recent
proof of the averaged null energy condition (ANEC) \cite{Maldacena:2015waa,Hartman:2016lgu}, which are reviewed in the next section.

This paper is organized as follows.  In section \ref{sec:review} we review the main ingredients regarding the operator product expansion, its convergence, and the ensuing positivity and boundedness properties of discontinuities in Lorentzian signature; we also
present a simplified dispersion relation, valid in the Regge limit, and discuss its relationship to the just mentioned recent work.
Section \ref{sec:fg} is purely mathematical and is devoted to deriving our main result, the inversion formula in eq.~(\ref{fg}).
The starting point will be the partial wave expansion in \cite{Costa:2012cb}, in which scaling dimensions are continuous,
and a corresponding not-so-well-known Euclidean inverse to this formula, which exploits the orthonormality of blocks.

In section \ref{sec:spin} we analyze the formula in the limit of large spin in a general conformal field theory,
substituting in the convergent OPE expansion in a cross-channel to re-derive and extend a number of results
from the analytic boostrap.
Section \ref{sec:gravity} discusses the simplifications in large-$N$ theories with a large gap,
and novel bounds on the contributions of ``heavy'' operators to the crossing relation,
with a brief discussion of loops in the bulk gravity theory. Section \ref{sec:conclusion} contains concluding remarks.
A lengthy appendix \ref{app:blocks} details formulas for handling conformal blocks in various dimensions,
while appendix \ref{app:Ising} details tests in the 2D Ising model.

\section{Review and main ingredients}\label{sec:review}

\subsection{Four-point correlator and conformal blocks}
\label{ssec:blocks}

We will be interested in the correlator of four conformal primary operators (which we will take to be scalars).
Up to an overall factor, it is a function of cross-ratios only:
\be
\langle \OO_4(x_4)\cdots\OO_1(x_1)\rangle = 
  \frac{1}{(x_{12}^2)^{\frac12(\Delta_1+\Delta_2)}(x_{34}^2)^{\frac12(\Delta_3+\Delta_4)}}
  \left(\frac{x_{14}^2}{x_{24}^2}\right)^{a}
  \left(\frac{x_{14}^2}{x_{13}^2}\right)^{b}
   \GG(z,\zbar) \label{stripped_G}
\ee
where here and below $a=\frac12(\Delta_2-\Delta_1)$, $b=\frac12(\Delta_{3}-\Delta_4)$,
and $z$, $\zbar$ are conformal cross-ratios
\be
 z\zbar=\frac{x_{12}^2x_{34}^2}{x_{13}^2x_{24}^2},\qquad (1-z)(1-\zbar) = \frac{x_{23}^2x_{14}^2}{x_{13}^2x_{24}^2}\,.
\ee
The operator product expansion (OPE) produces a series expansion around the limit where two points
coincide. The expansion in the $s$-channel (between 1 and 2) reads
\be
 \GG(z,\zbar) = \sum_{\j,\Delta} f_{12\OO}f_{43\OO}\,G_{\j,\Delta}(z,\bar{z})  \label{OPE}
\ee
where the sum runs over the spin $\j$ and dimension $\Delta$ of the exchanged primary operator $\OO$.
The conformal blocks $G$ are special functions which resum derivatives (``descendants") of $\OO$.
They are eigenfunctions of the quadratic and quartic Casimir invariants (\ref{Casimir_ops}) of the conformal group.
It will be useful to use blocks normalized so that, at small $z\ll \zbar$:
\begin{flalign}\label{normalization}
&& G_{\j,\Delta}(z,\zbar) &\to z^{\frac{\Delta-\j}{2}}\zbar^{\frac{\Delta+\j}{2}} &(0\ll z\ll \zbar \ll 1)\,.
\end{flalign}
The same normalization was used in \cite{Simmons-Duffin:2016wlq}.
The angular dependence when $z$ and $\zbar$ are both small but of comparable magnitude
can be expressed in terms of Gegenbauer polynomials, see eq.~(\ref{gegen}).
In even spacetime dimensions, the conformal blocks admit closed-form expressions in terms of hypergeometric functions,
for example
\begin{flalign}\label{blocks24}
&& G_{\j,\Delta}(z,\zbar) &= \frac{k_{\Delta-\j}(z)k_{\Delta+\j}(\zbar)+k_{\Delta+\j}(z)k_{\Delta-\j}(\zbar)}{1+\delta_{\j,0}}
 &(d=2)\,,
\\
&& G_{\j,\Delta}(z,\zbar) &= \frac{z\zbar}{\zbar-z}\big[
 k_{\Delta-\j-2}(z)k_{\Delta+\j}(\zbar)-k_{\Delta+\j}(z)k_{\Delta-\j-2}(\zbar)\big] & (d=4)\,.
\end{flalign}
In both expressions, $k_\beta$ denotes the hypergeometric function
\be
k_{\beta}(z) = z^{\beta/2}\,{}_2F_1(\beta/2+a,\beta/2+b,\beta,z)\,.
\ee

\begin{figure}
\be\begin{array}{cc}
\hspace{6mm}\def\svgwidth{80mm}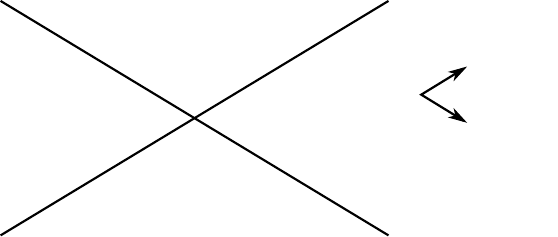
&
\hspace{6mm}\def\svgwidth{70mm}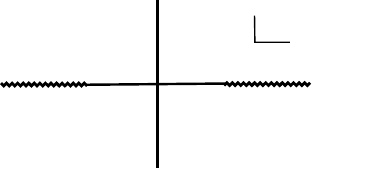
\\
\\ \hspace{-17mm}\mbox{(a)}&\hspace{-5mm}\mbox{(b)}
\end{array}\nonumber\ee
\caption{(a) Four points in Lorentzian signature, with time running upward.  The pairs $x_4-x_1$ and $x_2-x_3$ are timelike separated
for $\rhobar>1$.
(b) Corresponding configuration of cross-ratios $\rho$ and $\rhobar$.
}
\label{fig:rindler}
\end{figure}

Since four points can always be mapped to a plane via a conformal transformation,
$z$ and $\zbar$ can be viewed as coordinates on a two-dimensional plane.
In fact it will be convenient to parametrize the four points in a more symmetrical way,
following \cite{Hogervorst:2013sma}.  Since we'll be interested in Lorentzian kinematics,
we distribute the points symmetrically in two Rindler wedges.
In terms of lightcone coordinates $\rho,\rhobar=x^1\mp x^0$, we let, as shown in figure~\ref{fig:rindler}:
\be
  x_4=(\rho,\rhobar)=-x_3\,,\qquad  x_1=(1,1) = -x_2\,.
\ee
Their cross-ratio evaluates to $z=\frac{4\rho}{(1+\rho)^2}$, or, equivalently:
\be
 \rho = \frac{1-\sqrt{1-z}}{1+\sqrt{1-z}},\qquad
 \rhobar = \frac{1-\sqrt{1-\zbar}}{1+\sqrt{1-\zbar}}. \label{rhorhobar}
\ee
For convergence purposes, in the $\rho$-variables
the blocks behave essentially like power-laws: $G_{\j,\Delta}(z,\zbar)
\approx \rho^{\frac{\Delta\pm \j}{2}}\rhobar^{\frac{\Delta\mp \j}{2}}$.
In particular, the first singularities being at $\rho=\pm 1$,
the OPE converges whenever $\rho,\rhobar$ are both within the unit disc;
for rigorous estimates we refer to \cite{Pappadopulo:2012jk}.
Note that this holds whether or not $\rho$ and $\rhobar$ are complex conjugate of each other, which will be important below.

The unit $\rho$-disc covers the full complex $z$-plane, as can be seen from the fact that
$\rho$ and $1/\rho$ project onto the same value of $z$. They represent, however, physically distinct Lorentzian configurations as we now discuss.

\subsection{Positivity and analyticity properties of Rindler wedge correlator}

We will be interested in Lorentzian kinematics where the lightcone coordinates $\rho$ and $\rhobar$ are independent
real variables.  As long as both are positive, the four points lie within two disjoint Rindler (spacelike) wedges.
When both $\rho$ and $\rhobar$ are small, all points are spacelike separated and the physics is essentially Euclidean.
We will be more interested in the case $0<\rho<1<\rhobar$, where, as depicted in figure \ref{fig:rindler}(a), the distances $x_4-x_1$ and $x_2-x_3$ both become timelike.
According to the standard Feynman's $i0$ prescription ($x^0\to x^0(1-i0)$),
$\rhobar$ should then be slightly below the cut if we are computing the time-ordered correlator, and above the cut for its complex conjugate.

These kinematics do not lie within the radius of convergence of the $s$-channel sum (\ref{OPE}).
They lie, however, within the radius of convergence of the $t$-channel sum, which is based around the limit $x_2{\to}x_3$
corresponding to $\rho,\rhobar=1$.
This expansion is obtained by interchanging $z$ and $1-z$, which using (\ref{rhorhobar})
gives a somewhat nontrivial transformation for $\rho$:
\be
 \GG(\rho,\rhobar) = \frac{(z\zbar)^{\frac{\Delta_1+\Delta_2}{2}}}{\big((1{-}z)(1{-}\zbar)\big)^{\frac{\Delta_2+\Delta_3}{2}}}
 \sum_{\j',\Delta'}
\tilde f_{23\OO}\tilde f_{14\OO}
 \left(\frac{1-\sqrt{\rho}}{1+\sqrt{\rho}}\right)^{\Delta'+\j'}
 \left(\frac{1-\sqrt{\rhobar}}{1+\sqrt{\rhobar}}\right)^{\Delta'-\j'}\,. \label{t-channel_OPE}
\ee
We put the primes to remind ourselves that these quantum numbers are exchanged in the $t$-channel.
In this subsection, to simplify the argument (and with no loss of generality) we will not use the full conformal blocks but rather just
power laws, that is we include both primaries and descendants independently;
the notation $\tilde{f}$ reminds us of that.
The sum converges provided that the parentheses are within the unit disc, which is the case for $\rho$ and $\rhobar$ within
the complex plane minus the negative axis: $\rho,\rhobar\in C\setminus (-\infty,0]$.

Let us first consider the case where operators $1$ and $2$ are identical, and also $3$ and $4$.
The OPE coefficients $\tilde f_{23\OO}\tilde f_{14\OO}$, as squares of real numbers, are then positive.
What happens in the Lorentzian region $0{<}\rho{<}1{<}\rhobar$ is that the scaling blocks acquire a phase:
\be
\GG(\rho,\rhobar) = \big|\cdots\big|\sum_{\j',\Delta'} \tilde f_{23\OO}\tilde f_{14\OO} 
  \left|\frac{1-\sqrt{\rho}}{1+\sqrt{\rho}}\right|^{\Delta'+\j'}\left|\frac{1-1/\sqrt{\rhobar}}{1+1/\sqrt{\rhobar}}\right|^{\Delta'-\j'}
e^{i\pi(\Delta'-\j'-\Delta_2-\Delta_3)}, \label{t-channel_OPE1}
\ee
where the dots stand for the prefactor in (\ref{t-channel_OPE}).
Since the absolute value of each term is the same as for the positive sum corresponding to the Euclidean point $\rhobar\to 1/\rhobar$,
one thus find the following inequality:
\be
 \big| \GG(\rho,\rhobar)\big| \leq \GG(\rho,1/\rhobar)\equiv \GG_{\rm Eucl}(\rho,\rhobar) \qquad\qquad (0<\rho<1<\rhobar)\,. \label{inequality}
\ee
Intuitively, this states simply that the amplitude for a projectile crossing a target
is smaller than the amplitude for them to propagate independently.
This is analogous to flat space scattering,
where $S$-matrix elements between normalized states satisfy $|S|\leq 1$.
In this context, it is conventional to subtract off the free propagation by writing $S=1+i\mathcal{M}$,
and the imaginary part of the amplitude then satisfies ${\rm Im}\,\mathcal{M}\geq 0$.
The above inequality means that we can similarly define a CFT ``amplitude'' with a positive imaginary part:
\be
 i\mathcal{M}\equiv \GG(\rho,\rhobar)-\GG_{\rm Eucl}(\rho,\rhobar)\qquad\Rightarrow\qquad
 {\rm Im}\,\mathcal{M}\geq 0\,. \label{defM}
\ee
This imaginary part is equal to a \emph{double} discontinuity of the correlator:
\be\begin{aligned}
\label{dDisc}
{\rm Im}\,\mathcal{M}\equiv 
\dDisc \GG(\rho,\rhobar)&\equiv \GG_{\rm Eucl}(\rho,\rhobar)-\tfrac12\GG(\rho,\rhobar-i0)- \tfrac12\GG(\rho,\rhobar+i0)&
\\ &\geq 0 \hspace{60mm}(0<\rho<1<\rhobar).
\end{aligned}\ee
We will find below that ${\rm Im}\,\mathcal{M}$ is the argument of the CFT Froissart-Gribov formula.

It may seem unfamiliar that the imaginary part
is equal to a \emph{double} discontinuity (as opposed to a single discontinuity for the usual $S$-matrix)
but intuitively the role of the extra discontinuity is to subtract the ``1'' part of the $S$-matrix from the correlator.
This fact will be crucial below when we discuss large-$N$ theories.

Since the $i0$ prescription on the time argument
encodes the operator ordering, the double discontinuity can also be written as a commutator squared:
\be
\dDisc \GG(\rho,\rhobar) =  -\tfrac12\langle0| [\OO_2(-1),\OO_3(-\rho)][\OO_1(1),\OO_4(\rho)]|0\rangle \geq 0\,. \label{commutator}
\ee
Indeed one can check that for the two terms with ``non-scattering'' operator ordering, like
$\langle \OO_2(-1)\OO_3(-\rho)\OO_4(\rho)\OO_1(1)\rangle$, the continuation path for the cross-ratios $z,\zbar$ is equivalent
to staying within the Euclidean region, thereby reproducing the $\GG_{\rm Eucl}(\rho,\rhobar)$ term in eq.~(\ref{dDisc}).

Positivity of the commutator squared (\ref{commutator}) has appeared in several recent works.
It holds, in any QFT (not necessarily conformal), due to the Cauchy-Schwartz inequality
together with the property of so-called Rindler positivity (see refs.~\cite{Maldacena:2015waa,Hartman:2016lgu} and section 3 of \cite{Casini:2010bf}).
The argument here (similar to \cite{Hartman:2015lfa,Maldacena:2015iua}), valid in conformal field theories,
relied only on the usual positivity of Euclidean ($t$-channel) OPE coefficients.

For unequal operators, the more precise definition of the double discontinuity
is that it should be taken with respect to the
two time-like invariants $x_{14}^2$ and $x_{23}^2$ successively, at the level of the unstripped correlator on the left-hand-side of (\ref{stripped_G}).
That is, one should take the difference between the two different ways of making these invariants timelike,
$x_{jk}^2\to -|x_{jk}^2|\pm i0$.
This gives, when translated to the stripped correlator,
\be\begin{aligned}
 \dDisc \GG(\rho,\rhobar) &\equiv \cos(\pi(a+b)) \GG_{\rm Eucl}(\rho,\rhobar) 
\\& \qquad-\tfrac12e^{i\pi(a+b)} \GG(\rho,\rhobar-i0)
 -\tfrac12 e^{-i\pi(a+b)} \GG(\rho,\rhobar+i0)\,,
\label{dDisc_full}
\end{aligned}\ee
again in the range $0{<}\rho{<}1{<}\rhobar{<}\infty$.
As a function of the four operators, $ \dDisc \GG_{1234}$ defines a positive-definite matrix
with respect to the two pairs $12$ and $34$ (e.g., it is a positive number whenever $1$ and $2$ stand for
the same linear combination of operators, and similarly for $3$ and $4$).

\begin{figure}
\be\begin{array}{cc}
\hspace{10mm}\def\svgwidth{80mm}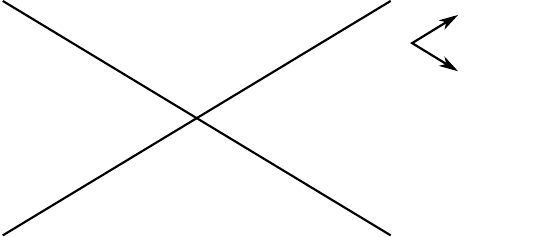
&
\hspace{5mm}\def\svgwidth{70mm}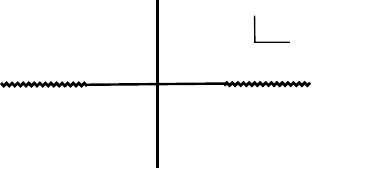
\\
\\ \hspace{-11mm}\mbox{(a)}&\hspace{-6mm}\mbox{(b)}
\end{array}\nonumber\ee
\caption{(a) Correlator in the Regge limit $\rho\propto 1/E\to 0$, $\rhobar\propto E\to \infty$
(with an overall boost applied to figure~\ref{fig:rindler} to make the figure more symmetrical).
(b) The crossing symmetry path $E\to Ee^{-i\pi}$, which interchanges points $3$ and $4$.
}
\label{fig:regge}
\end{figure}

Convergence of the $t$-channel OPE
(\ref{t-channel_OPE}) within the mentioned cut plane also implies an important analyticity property,
representing crossing symmetry.
Namely, starting from the timelike region $0{<}\rho{<}1{<}\rhobar$, one can rotate $\rho$ and $\rhobar$ by opposite phases
as shown in figure \ref{fig:regge}(b) and reach the region $\rhobar{<}{-}1{<}\rho{<}0$. This represents crossing, since,
in the parametrization (\ref{rhorhobar}), $\rho \to -\rho$ is the same as interchanging operators $3$ and $4$.
This crossing path is reminiscent of the Epstein-Glaser-Bros path in axiomatic $S$-matrix theory \cite{Bros:1965kbd},
and, just like in that context, it lands us on the ``wrong'' side of the cut, e.g. on the complex conjugate (anti-time-ordered) amplitude.
Note that for this crossing path $\rho$ and $\rhobar$ must be rotated in opposite directions,
because the Euclidean correlator near $\rho,\rhobar=0$ is single-valued only when $\rho$ and $\rhobar$ are complex conjugate of each other.

If one were to rotate $\rho$ and $\rhobar$ in the same direction, one would land in a physically completely different
kinematics, where the four points are inside timelike Milne wedges instead of the spacelike Rindler wedges.
This region is relevant to bulk high-energy scattering in AdS/CFT and contains the so-called ``bulk point'' limit in two dimensions (see \cite{Heemskerk:2009pn,Maldacena:2015iua}).  Although this region is interesting, we will not directly use it in this paper.

\subsection{Toy dispersion relation, ANEC and the bound on chaos}
\label{ssec:toy disp}

The above analyticity and boundedness properties immediately imply
a simple-minded dispersion relation, whose integrand is positive definite at least in
the Regge limit (large boost acting on 1 and 2).
In this subsection we will parametrize this limit as $E\to\infty$ with
\be
 \rho= \sigma/E, \qquad \rhobar= E \,.
\ee
With no loss of generality (because of the symmetry between $z$ and $\zbar$) we will assume that $\sigma<1$.

A key fact is that the correlator is bounded and approaches a limit as $E\to\infty$ (with $\sigma$ fixed):
this is because the $t$-channel OPE (\ref{t-channel_OPE}) is dominated by its Euclidean counterpart which converges to $\GG(0,0)=1$.
For operators that are not identical, the correlator is also bounded since the individual OPE coefficients satisfy
$\big|\tilde{f}_{23\OO}\tilde{f}_{14\OO}\big|< {\rm max}(\tilde{f}_{23\OO}^{\,2},\tilde{f}_{14\OO}^{\,2})$.

In fact, the correlator approaches a finite constant as $|E|\to\infty$ along any complex direction in the lower-half plane.
This can be shown by combining the $t$-channel and $u$-channel OPE.
By itself, the $t$-channel OPE only gives a numerically
weaker bound $|\GG|<1/\cos((\arg E)/2)^{\Delta_1+\Delta_2}$
since only the real part of $\sqrt{\rho}$ damps the exponent for each term.
This bound becomes poor near ${\rm arg}\,E=-\pi$, but in this case one can use the $u$-channel OPE instead, which converges nicely.
Being bounded like this, analyticity then implies that it approaches the same constant not only along the real axis,
but also along any complex direction $-\pi\leq{\rm arg}\,E\leq 0$:
\be
 \lim_{|E|\to\infty}\mathcal{M}(\sigma,E) = C.
\ee
This can be proved easily by taking a derivative of the contour integral (\ref{contour1}) below, dropping the arc at infinity,
and integrating back.
Under crossing $t\leftrightarrow u$, $C$ goes to $-C^*$.\footnote{
The statement that ${\rm Im}\,C$ is crossing symmetric is essentially Pomeranchuk's theorem,
which states that proton-proton and proton-antiproton total cross-sections are asymptotically equal if they grow with energy.}

To obtain a dispersion relation for $\mathcal{M}(\sigma,E)$, we simply write down the contour integral
\be
 \mathcal{M}(\sigma,E) = \frac{1}{2\pi i}\oint_{C} \frac{dE'\,\mathcal{M}(\sigma,E')}{E-E'}\,, \label{contour1}
\ee
where $E$ is assumed to be in the lower-half plane and the contour encircles the lower-half plane clockwise.
In fact it doesn't hurt to add ``0'' in the form of a similar integral but encircling the upper-half plane,
with the integrand replaced by the  analytic function which is equal to $\mathcal{M}^*$ just above the real axis.
The half-circles at infinity then simply add up to the real part of $C$:
\be
 \mathcal{M}(\sigma,E) = {\rm Re}\,C + \frac{1}{\pi} \int_{-\infty}^\infty \frac{dE'\,{\rm Im}\,\mathcal{M}(\sigma,E')}{E-E'}. \label{toy_dispersion}
\ee
(Note that the integral diverges logarithmically if ${\rm Im}\,C\neq 0$, however with opposite signs at $E'\to \pm\infty$
and such that it converges to the correct result if symmetrized under $E'\to -E'$ under the integration sign.)

The ``toy'' dispersion relation (\ref{toy_dispersion}) converges to the four-point correlator but a caveat
is that for ``low'' energies $|E'|< 1$ one leaves the Lorentzian region and the integrand switches sign,
since going below $\rhobar=1$ interchanges $\mathcal{M}$ and $\mathcal{M}^*$
according to the definition (\ref{defM}).
Furthermore, when $|E'|<\sigma$, one loses control over the sign and even reality of the integrand.
However, for $|E|\gg 1$, these region contribute only a subdominant amount.
This is the sense in which the above is only a toy dispersion relation, but for this reason
the subsequent discussion should be understood to hold only for high enough energies $|E| \gg 1$.

The toy dispersion relation still contains interesting physics. Separating explicitly the real and imaginary parts of $E$: $E=x-iy$,
the following inequalities follow from straightforward differentiation, assuming only that ${\rm Im}\,\mathcal{M}>0$ on the real axis as shown above:
\be\begin{aligned}
 {\rm Im}\,\mathcal{M}(x-iy,\sigma) &= \frac{y}{2\pi} \int_{-\infty}^\infty \frac{dE'\,2{\rm Im}\,\mathcal{M}(E',\sigma)}{(E'-x)^2+y^2}
 \\ & \geq0,
\\
(y\partial_y -1) \log
{\rm Im}\,\mathcal{M}(x-iy,\sigma) &=
-\frac
{\int_{-\infty}^\infty dE'\,2{\rm Im}\,\mathcal{M}(\sigma,E')\frac{y^2}{((E'-x)^2+y^2)^2}}
{\int_{-\infty}^\infty dE'\,2{\rm Im}\,\mathcal{M}(\sigma,E')\frac{1}{(E'-x)^2+y^2}}
\\ & \leq0 . \label{inequalities}
\end{aligned}\ee
The first of these inequality, which essentially states that $|S|\leq 1$ holds throughout the complex lower half-plane
and not only the real axis, played an important role in a recent proof of the averaged
null energy condition (ANEC) \cite{Hartman:2016lgu} (see also \cite{Hofman:2016awc}).
There one showed using the lightcone OPE that there is a regime where the left-hand-side is dominated
by exchange of the stress tensor integrated over a null line, thereby proving positivity of its matrix elements,
and also generalized this argument to the operator of lowest twist for each even spin $\geq 2$.

The second inequality states that, locally, the amplitude cannot grow faster than linearly (along the imaginary axis).
When expressed in terms of Rindler time $t=\log E$ and temperature $T=1/(2\pi)$,
this is equivalent to the ``bound on chaos'' on the Lyapunov exponent, $\lambda \leq 2\pi T$, proved in ref.~\cite{Maldacena:2015waa}.
(As discussed in appendix A of \cite{Maldacena:2015waa}, the present context of high-energy CFT scattering can be viewed as a special case of their
general finite-temperature results applied to Rindler space. Specific properties of CFTs, tied to OPE convergence,
apparently allowed us here to simplify some steps of the proof,
for example the Phragmen-Lindelof argument used there.)

Although the bound on chaos will not be used directly in this paper, the closely related dispersion relation
(\ref{toy_dispersion}) is conceptually central and our main goal will be to extend it to OPE coefficients.

\subsection{Invitation: theories with large $N$ and large gap}\label{ssec:largeN}

The toy dispersive representation (\ref{toy_dispersion}) tells us more.
For example, to leading order in a large-$N$ theory, the double discontinuity $\rm Im\,\mathcal{M}$
has the neat property that it is insensitive to double-trace operators exchanged in the cross channels.
This can be seen again from the $t$-channel OPE (\ref{t-channel_OPE1}), because double-trace operators
have dimensions $\Delta_2+\Delta_3$ or $\Delta_1+\Delta_4$ plus integers,
and the resulting continuation phases are then killed by the double discontinuity (\ref{dDisc_full}) in both cases.

In the case of identical operators,
the double-trace OPE coefficients are enhanced by a factor $N^2$ compared with the connected contribution since they contribute
at the disconnected level.
However, the connected contribution, associated with the leading $1/N^2$ corrections to coefficients and anomalous dimensions,
contains at most a single logarithm of $(1-\zbar)$ (see for example \cite{Heemskerk:2009pn}).
These double-trace contributions are thus again killed by the double discontinuity.

The fact that the imaginary part (\ref{dDisc_full}) is sensitive only to single-trace operators
is consistent with intuition from gauge-gravity correspondence: the imaginary part arises from
``on-shell'' exchanged states, and in tree-level AdS/CFT these are the elementary bulk fields dual to single-trace operators.
Thus the toy dispersive representation (\ref{toy_dispersion}) tells us physically that the double-trace information
is fully controlled by what the single-traces do (strictly speaking, at this point, up to the neglected $|E'|<\sigma$ contributions).

Consider specifically the case where the single-trace spectrum is sparse, in the sense
that the lowest twist operator of spin higher than 2 has a large twist $\Delta_{\rm gap}\gg 1$.
The spectral density will then be suppressed, as the contribution of heavy operators to the $t$-channel OPE has the form
\be \label{heavy_contribution}
\dDisc \GG(z,\zbar)\sim
\sum_{\Delta\geq \Delta_{\rm gap}} \left(\frac{(1-\sqrt{\rho})(1-1/\sqrt{\rhobar})}{(1+\sqrt{\rho})(1+1/\sqrt{\rhobar})}\right)^{\Delta}\propto 
 e^{-2\Delta_{\rm gap}(\sqrt{z}+\sqrt{\zbar})} \sim e^{-\Delta_{\rm gap}/\sqrt{E}}.
\ee
Thus the ``energy'' variable $E$ introduced above is indeed essentially equivalent to CFT energies, e.g. scaling dimensions (squared),
and heavy operators produce an imaginary part only at high energies, again in line with AdS/CFT intuition.

Plugging this into the second inequality (\ref{inequalities}) one immediately sees that any theory with a sparse spectrum
will nearly saturate the bound on chaos within at least the energy range $1\ll E\ll \Delta_{\rm gap}^2$
(provided only that the light operator contribution to the dispersive relation is not enhanced by a power of $\Delta_{\rm gap}^2$,
which would preclude the heavy operators from dominating it.)
Conversely, the bound on chaos can only be saturated, locally for some value of energy,
if most of the spectral density is concentrated at a much higher energy, which is a weak statement of a ``gap.''
Also, since ${\rm Im}\,\mathcal{M}$ is locally bounded, the gap certainly can't be larger than the inverse coefficient of the linear growth,
which is essentially the stress tensor two-point function $c_T$; this reflects the familiar observation that in weakly coupled gravity theories, the string scale never exceeds the Planck scale.

These are all moral implications of the dispersion relation, but to make them (and others) fully quantitative
one would like to have a dispersion relation whose integrand remains physically sensible and positive
even away from the Regge limit. We were not able to derive one.
However, in the next section we will short-circuit this technical issue by switching our focus to the OPE data
instead of the correlator itself, and deriving a \emph{Froissart-Gribov formula}.

\subsection{From dispersion relation to Froissart-Gribov formula}
\label{ssec:fg0}

The importance of the concept of analyticity in spin was emphasized already in the introduction,
for example in the two contexts of large spin expansions and also bulk locality.
Being related to analyticity in energy, itself tied to causality, in a sense it is a physical reflection of causality.
The Froissart-Gribov formula is an integral representation for partial wave coefficients
which makes analyticity manifest and quantitative.
Here we briefly review its connection to dispersion relation in the context of the flat-space $S$-matrix.
One considers $2\to 2$ scattering, and define as usual the projection of the amplitude
into the partial wave of angular momentum $\j$:
\be
 a_\j(s) = \int_{-1}^1 d(\cos\theta) (\sin\theta)^{d-4}\,C_\j(\cos\theta) \mathcal{M}(s,t(\theta)),\qquad
 t(\theta)=-\tfrac{s-4m^2}{2}(1-\cos\theta), \label{flat_space_eucl}
\ee
where $C_\j(\cos\theta)$ are Gegenbauer polynomials (Legendre polynomials $P_\j$ in four spacetime dimensions).
Here $s,t,u$ are the usual Mandelstam variables for $2\to 2$ scattering.
The Froissart-Gribov formula follows from using a fixed-$s$ dispersive representation of the amplitude,
in terms of $t$- and $u$-channel cuts:
\be
 \mathcal{M}(s,t) =
   \int_{t_0}^\infty \frac{dt'}{t'-t-i0} {\rm Disc}_t\,\mathcal{M}(s,t') + (t\leftrightarrow u)\,. \label{flat_space_disp}
\ee
The integration threshold $t_0$ will not be important, and convergence and subtractions will be discussed shortly.
To see analyticity in spin, one simply plugs eq.~(\ref{flat_space_disp}) into (\ref{flat_space_eucl});
changing variable to $t'=\tfrac{s-4m^2}{2}(\cosh\eta-1)$ this gives:
\be
 a_\j(s) = \int_{-1}^1 dx' (1-x'^2)^{\frac{d-4}{2}}\,C_\j(x')
 \int_{\eta_0}^\infty \frac{d(\cosh\eta)}{\cosh\eta-x'} {\rm Disc}_t\,\mathcal{M}(s,t') + (t\leftrightarrow u)\,.
\ee
Interchanging the order of integrations, the $x'$-integral can be done once and for all.
Defining
\be
 Q_\j(x) \equiv \int_{-1}^1 \frac{dx'\,C_j(x')}{x-x'} \left( \frac{1-x'^2}{1-x^2}\right)^{\frac{d-4}{2}}\,,
\ee
one thus find:
\be
 a_\j(s) = a_\j^t(s) + (-1)^\j a_\j^u(s), \label{flat_tu}
\ee
where
\be\begin{aligned}
 a_\j(s) &= \int_{1}^\infty d(\cosh\eta)\,(\sinh\eta)^{d-4} Q_\j(\cosh\eta)\,{\rm Disc}_t\,\mathcal{M}(s,t),\qquad t=\tfrac{s-4m^2}{2}(\cosh\eta-1)\,. \label{FG_flat}
\end{aligned}\ee
Eq.~(\ref{FG_flat}) is known as the Froissart-Gribov formula.  It shows that, while $a_\j$ was a-priori defined only for integer $\j$,
its $t$- and $u$-channel contributions $a_j^{t,u}(s)$ are separately analytic in spin.
They are power-behaved at large imaginary $\j$, as opposed to the direct evaluation of (\ref{flat_space_eucl}) which would
grow like $e^{\pm i\pi \j}$, and general arguments show that an analytic continuation with this property is unique.\footnote{
Because of the $(-1)^\j$ multiplying the $u$-channel cut contribution,
the even and odd spin partial waves constitute two independent well-behaved analytic functions.}

The regime of validity of the Froissart-Gribov formula (\ref{FG_flat}) is the same as that of the dispersion relation (\ref{flat_space_disp}) from which is originates.
Depending on the high-energy behavior, this dispersion relations may receive ambiguities, known as subtractions,
that are polynomials in $t$.  These polynomials affect a finite number of the $a_\j$'s in eq.~(\ref{flat_space_eucl}).
Thus partial wave coefficients are analytic in spin \emph{except} for a finite number of low spins, which
cannot be obtained from the Froissart-Gribov integral.
The minimum spin starting from which the formula works is equal to the exponent controlling the power behavior of the amplitude
in the Regge limit (large $|s|$ at fixed $t$).

In the next section, we will derive using group-theoretic arguments a Froissart-Gribov formula for OPE data
in conformal field theories. The derivation will by-pass the dispersive representation which we do not have, but to which the formula
is morally equivalent. The good high-energy behavior of CFT correlators discussed above, will imply that the formula works for all spins except possibly $\j=0,1$.

\section{Inverting the OPE: The CFT Froissart-Gribov formula}\label{sec:fg}

A prerequisite first step to derive integral representations is to get rid of the discreteness
of $\Delta$.  This first step was already carried out in ref.~\cite{Costa:2012cb}, where the sum over dimensions
was replaced by an integral over continuous dimensions:
\be
 \GG(z,\zbar) = 1_{12}1_{34}+
 \sum_{\j=0}^\infty \int_{d/2-i\infty}^{d/2+i\infty} \frac{d\Delta}{2\pi i} \,\c(\j,\Delta)\,F_{\j,\Delta}(z,\zbar). \label{eucl_expansion}
\ee
The first term is the identity contribution.
The goal of this section is to derive a Lorentzian inverse to this representation, which will express the coefficients $\c(\j,\Delta)$
\emph{analytically in $\j$} and \emph{in terms of positive Lorentzian data}, in analogy with the Froissart-Gribov formula (\ref{FG_flat}).

This section is exclusively mathematical. After briefly reviewing the expansion (\ref{eucl_expansion}),
we derive an analog to the Euclidean inverse (\ref{flat_space_eucl}) and then perform its analytic continuation 
to Lorentzian signature, obtaining our main result (\ref{fg}).  Many technical details are moved to appendix \ref{app:blocks}.
Physical implications are discussed in the next sections.

\subsection{Partial waves: Euclidean case}
\label{ssec:eucl}

The idea behind eq.~(\ref{eucl_expansion}) is to expand correlators over an orthogonal basis of eigenfunctions
$F_{\j,\Delta}$ of the Casimir invariants of the conformal group \cite{Costa:2012cb}.
These are the quadratic and quartic differential operators in eqs.~(\ref{Casimir_ops}).

These invariants are self-adjoint only when acting on single-valued complex functions (that is, which do not have branch cuts in Euclidean kinematics $\zbar=z^*$), otherwise, integration-by-parts would receive extra boundary terms from the branch cuts.
Thus such an expansion only has a chance to work in the space of single-valued functions.
Fortunately, the physical correlator $\GG(z,\zbar)$ is of this type.
The individual conformal blocks $G_{\j,\Delta}$ entering the OPE sum are not, and
to make a basis of eigenfunctions one must use instead the single-valued combinations \cite{Costa:2012cb}, also known as harmonic functions:
\be
 F_{\j,\Delta}(z,\zbar) = \frac12\left(G_{\j,\Delta}(z,\zbar) + \frac{K_{\j,d-\Delta}}{K_{\j,\Delta}} G_{\j,d-\Delta}(z,\zbar)\right), \label{F_single}
\ee
whose coefficients can be expressed using a frequently-recurring products of $\Gamma$-function:
\be
K_{\j,\Delta} = \frac{\Gamma(\Delta-1)}{\Gamma\big(\Delta-\tfrac{d}{2}\big)}\kappa_{\j+\Delta},
\qquad
\kappa_\beta=\frac{\Gamma\big(\tfrac{\beta}{2}-a\big)\Gamma\big(\tfrac{\beta}{2}+a\big)\Gamma\big(\tfrac{\beta}{2}-b\big)\Gamma\big(\tfrac{\beta}{2}+b\big)}
 {2\pi^2\Gamma(\beta-1)\Gamma(\beta)}\,. \label{kappa}
\ee
Single-valuedness of the harmonic functions $F_{\j,\Delta}(z,\zbar)$ can be understood from an integral representation \cite{SimmonsDuffin:2012uy},
which involves three-point functions of the exchanged operator and of its ``shadow'' related by $\Delta\to d-\Delta$.
This explains also why the shadow block appears in eq.~(\ref{F_single}).
With no loss of generality we can thus assume that shadow coefficients are symmetrical:
\be
\frac{\c(\j,\Delta)}{K_{\j,\Delta}}= \frac{\c(\j,d-\Delta)}{K_{\j,d-\Delta}}\,. \label{shadow}
\ee
For our purposes, single-valuedness of eq.~(\ref{F_single}) can also be verified explicitly using the analytic continuation formulas in appendix \ref{app:cont}.
It requires the spin to be an integer.

Under the assumption that the harmonic functions $F_{\j,\Delta}$'s are orthogonal, the expansion (\ref{eucl_expansion})
can be immediately inverted to give the $\c(\j,\Delta)$'s, by simply integrating against $F_{\j,\Delta}$:\footnote{I thank V.~Goncalves and J.~Penedones for initial collaboration on
this Euclidean inversion formula.  I have also been made aware of related work by B.~ van Rees and M.~Hogervorst
on this subject.}
\be
 \c(\j,\Delta) = N(\j,\Delta)
  \int d^2z\, \mu(z,\zbar)\, F_{\j,\Delta}(z,\zbar)\,\GG(z,\zbar)\,. \label{eucl_inversion}
\ee
The integration runs over the full complex plane (with $\zbar=z^*$).
The measure is fixed by self-adjointness of the Casimir differential operators in eq.~(\ref{Casimir_ops}), which works out
to give
\be
 \mu(z,\zbar) = \left|\frac{z-\zbar}{z\zbar}\right|^{d-2} \frac{\big((1-z)(1-\zbar)\big)^{a+b}}{(z\zbar)^2}\,. \label{measure}
\ee
The orthogonality assumption is actually a simple consequence of
the Casimir equations (modulo minor convergence issues to be discussed shortly).
The normalization factor $N(\j,\Delta)$, given in eq.~(\ref{eucl_normalization}),
can be calculated using only the behavior of the blocks near the origin,
where the radial integral boils down to a Mellin transform (which also explains the choice of contour in eq.~(\ref{eucl_expansion})).
Both of these issues are discussed in detail in \ref{app:ortho}.

We will not dwell too much on eq.~(\ref{eucl_inversion}), since it is only an intermediate step toward the
Lorentzian formula that we are after. Let us only describe the connection between the partial wave expansion
(\ref{eucl_expansion}) and the usual OPE sum (\ref{OPE}).
In Euclidean kinematics where $|\rho|<1$, the blocks $G_{\j,\Delta}(z,\zbar)$ vanish exponentially at large real $\Delta$,
like power-laws $|\rho|^{\Delta}$.
Therefore it is natural to close the contour to the right in eq.~(\ref{eucl_expansion})
on the first term of eq.~(\ref{F_single}), and to the left in the shadow block
(which will produce an identical result due to the mentioned shadow symmetry).
The OPE sum is therefore reproduced provided that the partial
wave coefficients $\c(\j,\Delta)$ have poles with appropriate residues:
\begin{flalign}
&&  c_{\j,\Delta}&=-\Res_{\Delta'=\Delta}\c(\j,\Delta')& (\mbox{$\Delta$ generic})\,. \label{c_generic}
\end{flalign}
The function $\c(\j,\Delta)$ thus encodes the usual OPE data
through the position and residue of its poles.  The fact that this data is encoded in an analytic function of $\Delta$ will be important physically,
since the discreteness of the original OPE data would otherwise preclude our next step.
The qualifier ``generic'' was added in the preceding equation due to some special cases that we now discuss.

\subsubsection*{Subtleties, convergence, contour, etc.}

This subsubsection contains technical comments that the reader may skip on first reading.
The preceding equations are too hasty for two reasons: the blocks $G_{\j,\Delta}$ themselves
have poles, and the integral (\ref{eucl_inversion}) doesn't always converge.
These two issues are fixed here.

Let us first discuss convergence of the inverse transform (\ref{eucl_inversion}) near $z=0$.
The integral in this region can be defined easily by analytic continuation.
Indeed one can make the integral convergent by subtracting finitely many terms in the small-$z$ expansion of the amplitude times block, which can then be integrated back using the analytic formula:
\be
 \int_0^1 \frac{d|z|}{|z|} |z|^{p\pm \Delta} \equiv \frac{1}{p\pm\Delta}\,. \label{simple_radial_integral}
\ee
Being analytic in $\Delta$, this prescription automatically preserves self-adjointness of the Casimir operators.
This formula also shows explicitly how the blocks in the usual OPE sum turn into poles of the function $\c(\j,\Delta)$.

There are a few special cases at low dimensions, which can all be understood by thinking of the 
representation (\ref{eucl_expansion}) as an inverse Mellin transform, which it is near the origin.
Due to the unitarity bound $\Delta-\j\geq d-2$ for $\j\geq 1$,
these only affect the $\j=0$ contribution.
\begin{itemize}
\item For operators with dimension less than $d/2$, one should deform the contour so it
picks only the positive-residue pole on the left of $d/2$ instead of its reflection on the right.
\item For operators of dimension precisely $d/2$, one should use a principal-value contour so that only half of the residue contributes.
(In two dimensions, this case also includes conserved currents with $\j=1$.)
\item The unit block is orthogonal to all $F_{\j,\Delta}$. This is why it appears separately in eq.~(\ref{eucl_expansion}).
\end{itemize}

The blocks $G_{\j,\Delta}$ have poles, which were not included in eq.~(\ref{c_generic}).
These poles are all below the unitarity bound $\Delta=\j{+}d{-}2$, but are nonetheless to the right of the contour $\Delta=d/2$,
and so they should be included.
Poles of conformal blocks were studied e.g. in \cite{Kos:2013tga,Kos:2014bka}.
Their possible location is heavily constrained by the Casimir equations, which imply that the residue solves
the same equations and thus must be related by one of the symmetries (\ref{symmetries}).
The only poles with $\Delta>d/2$ are at $\Delta=j+d-2-m$ with $m=0,1,2,\ldots$
and with residue proportional to $G_{j-1-m,j+d-1}$,
which is then a physical block (above the unitarity bound and with integer spin).
The proportionality factor $r_{\j,\Delta}$ is given in eq.~(\ref{r}).
Therefore all the residues produce physical blocks.
Collecting their coefficient, we find that the correct formula for the OPE coefficients is not (\ref{c_generic}) but rather
\be
 \c_{\j,\Delta} = -\Res_{\Delta'=\Delta}\left\{
 \begin{array}{lr}
 \c(\j,\Delta') & \mbox{($\Delta$ generic)}\,,\\
 \c(\j,\Delta') - r_{\j,\Delta'} \c(\Delta'{+}1{-}d,\j{+}d{-}1)&\qquad (\Delta{-}\j{-}d=0,1,2,\ldots)\,. \label{C_residues}
\end{array}\right.
\ee
This is consistent with the inversion formula (\ref{eucl_inversion})
since its integrand (through the factor $N(\j,\Delta)$) diverges at
$\Delta{-}\j{-}d=0,1,2$, producing poles unrelated to $z\to 0$
divergences. However these poles of the integrand cancels in the combination (\ref{C_residues}).

Finally let us analyze the region $z\to 1$ ($z\to\infty$ is similar).
Convergence there depends on the dimension of the lightest operator exchanged in the $t$-channel,
compared to that of the external operators. Thus for generic external operators the integral will not converge.
However one can define the integral by cutting off a small circle around 1, for example,
and dropping the singular terms as its radius goes to zero.
As long as the same cutoff is used for all spins (so that the
spurious pole cancelation just mentioned continues to operate), the resulting expression will only have the correct physical poles,
and the OPE coefficients extracted from eq.~(\ref{C_residues}) will correctly reproduce the correlator.
We conclude that the integral (\ref{eucl_inversion}) may or may not converge near $z\to1$ but this seems devoid of consequences
for generic $z,\zbar$. This is akin to the Fourier transform of a singular distribution, which is ambiguous up to contact terms
in coordinate space and polynomials in momentum space.

All of these subtleties can be tested explicitly in simple examples,
like the 2D Ising model, as discussed in appendix \ref{app:Ising}.

\subsection{$S$-matrix Froissart-Gribov formula revisited}

\begin{figure}
\be\def\svgwidth{80mm} 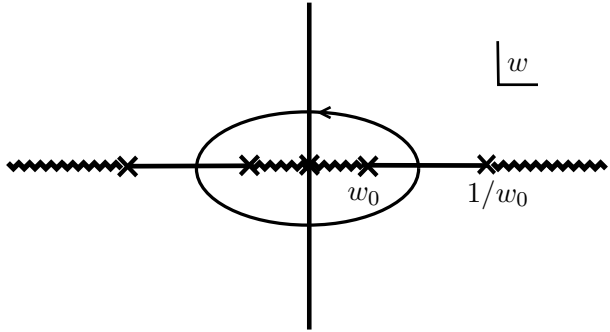 \nonumber\ee
\caption{Scattering amplitudes in the complex $w$-plane, where $w=e^{i\theta}$.
Two copies of the $t$-channel appear on the right, related to each other by $w\to 1/w$, and two copies of the $u$-channel cut appear on the left.
}
\label{fig:wplane}
\end{figure}

In section \ref{ssec:fg0} we saw how the angular momentum projection (\ref{flat_space_eucl})
and dispersion relation (\ref{flat_space_disp}) naturally combine into the ($S$-matrix)
Froissart-Gribov formula (\ref{FG_flat}), thus relating partial waves coefficients to discontinuities of amplitudes.
In CFT we have constructed analogous ingredients: the Euclidean inversion formula (\ref{eucl_inversion})
and the Regge-limit dispersion relation (\ref{toy_dispersion}), and one could imagine again substituting the later into the former.
Unfortunately, being restricted to the Regge limit, that dispersion relation is not good enough for our purposes.

Fortunately, there exists a second derivation of the $S$-matrix Froissart-Gribov formula, which does not require a dispersion relation.
It is essentially the contour deformation argument from the introduction;
we reformulate it here in terms of variables that will be more convenient shortly,
focusing for simplicity on $d=3$ scattering where the SO(2) Gegenbauer polynomials are just cosines.
The trick is to rewrite eq.~(\ref{flat_space_eucl}) as a contour integral in the variable $w=e^{i\theta}$:
\be
 a_\j(s) = \frac12\oint \frac{dw}{iw} \left(w^\j+w^{-\j}\right)\,\mathcal{M}(s,t(w)),\qquad t(w)= \frac{s-4m^2}{4}\left(w+\tfrac{1}{w}-2\right)\,. \label{eucl_contour}
\ee
The main features in the $w$-plane are shown in figure \ref{fig:wplane}.
On the positive real axis there are two branch points, where $t(w_0)=t_0>0$.
They are related by $w\to 1/w$ and both represent the same physical point, the $t$-channel threshold.
Similarly on the negative axis there are two copies of the $u$-channel cut.
To proceed, assuming $\j$ large enough that $w^\j$ times the amplitude vanishes near the origin of the unit disc,
for the first term we simply close the contour toward the interior.  There we find two branch cuts, corresponding to the $t$- and $u$-channels.
For the $w^{-\j}$ term we close outside but this gives the same result.
The integral becomes thus becomes a sum of contributions
from two cuts as in eq.~(\ref{flat_tu}):
\be
 a_\j^t(s) = \int_0^{w_0} \frac{dw}{w}w^\j\, {\rm Disc}_t\,\mathcal{M}(s,t(w))\,, \label{eucl_cut_integral}
\ee
which is equivalent to eq.~(\ref{FG_flat}) via the change of variable $2\cosh\eta=w+1/w$.
Whereas the Euclidean contour integral (\ref{eucl_contour}) only made physical sense for integer spin $\j$,
the discontinuity integral eq.~(\ref{eucl_cut_integral}) is fundamentally Lorentzian and makes sense for continuous
spin.

\subsection{Main derivation: conformal Froissart-Gribov formula}

\begin{figure}
\be\begin{array}{cc}
\def\svgwidth{65mm}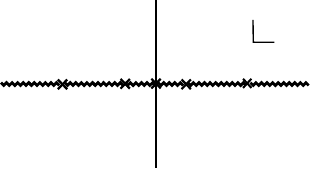
&\hspace{10mm}
\def\svgwidth{65mm}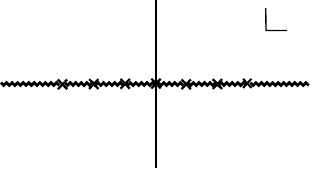
\end{array}\nonumber\ee
\caption{The $w$-plane contours $C$ and $C_\pm$ in the CFT case.  The singularities at $w=\pm 1$ are integrable and come
only from the measure factor $|z-\zbar|^{d-2}$.  The branch points at $w=\pm\sigma$ and $w=\pm1/\sigma$ pose no problems,
but the arcs at $w=0$ and $w\to\infty$ correspond to the Regge limit where the integrand should vanish at large positive spin.
}
\label{fig:wplane2}
\end{figure}

We now adapt this contour deformation argument to CFTs.
The geometry is easier to visualize in terms of the $\rho$-coordinates (\ref{rhorhobar}),
where the ``radial'' and ``angular'' variables are naturally identified with the magnitude and phase of $\rho$.
The Euclidean inversion formula (\ref{eucl_inversion}) then runs over the unit disc:
\be
  \c(\j,\Delta) = N(\j,\Delta) \int_0^1 \sigma d\sigma \oint_{|w|=1} \frac{dw}{i w} \,\mu(\rho,\rhobar)\, g(\rho,\rhobar)\, F_{\j,\Delta}(\rho,\rhobar)
  \Big|_{\begin{subarray}{l}\rho=\sigma w \\ \rhobar=\sigma w^{-1}\end{subarray}}
  \,. \label{eucl_inversion_rho}
\ee
Here, the measure in $\rho$-coordinates, including the Jacobian from the change of variable, is
\be
\mu(\rho,\rhobar)=\frac{(1-\rho^2)(1-\rhobar^2)}{16\rho^2\rhobar^2}
\left|\frac{(1-\rho\rhobar)(\rhobar-\rho)}{4\rho\rhobar}\right|^{d-2}\left(\frac{(1-\rho)(1-\rhobar)}{(1+\rho)(1+\rhobar)}\right)^{2(a+b)}\,. \label{measrho}
\ee
We also recall that, at this stage, $\j$ is still restricted to be an integer.

At fixed $\sigma<1$, the complex-$w$ plane contains the same features as in the preceding subsection.
There are two copies of the $t$-channel cut starting at $w=\sigma$ and $w=1/\sigma$ (corresponding to $\rhobar=1$ and $\rho=1$, respectively),
and two copies of the $u$-channel cut, starting at $w=-\sigma$ and $w=-1/\sigma$.

Roughly speaking, the idea is to split the block $F_{\j,\Delta}$ into a part which vanishes like $w^\j$ near the origin and another which vanishes
like $w^{-\j}$ at infinity, so that we can close the $w$-contour to the interior or exterior in each case.
This was the trick in the $S$-matrix case.
The behavior of the blocks $F_{\j,\Delta}$ is however more complicated.
As explained in appendix \ref{app:blocks}, there are 8 solutions to the Casimir equations corresponding to a given value of $\j$ and $\Delta$,
which can be conveniently labelled (for generic $\j$ and $\Delta$) by their power-law behavior in the regime $0\ll z \ll \zbar \ll 1$:
\be
 \gpure_{j,\Delta}(z,\zbar) = z^{\frac{\Delta-j}{2}}\zbar^{\frac{\Delta+j}{2}}\times (1 + \mbox{integer powers of $z/\zbar,\zbar$})\,. \label{gpure_text}
\ee
The 8 solutions are then related by the symmetries (\ref{symmetries}).
The limit $w\to 0$ is governed by these asymptotics but after analytic continuation around the point $\zbar=1$.
The block $F_{\j,\Delta}(z,\zbar)$ then becomes a complicated combination of all 8 basic solutions.
The possible exponents are such that two solutions vanish like $w^{\j}$, two diverge like $w^{-\j}$,
and the four remaining ones have exponents controlled by $\Delta$.
Ideally we would like to get rid of the 6 nondecreasing solutions, which is not obviously possible.
A systematic strategy is to simply close the contour in eq.~(\ref{eucl_inversion_rho}) to the interior of the unit disc,
and then try to remove the nonvanishing terms near the origin by adding zero
in the form of an integral over blocks $e^\pm_{\j,\Delta}(\rho,\rhobar)$ along the closed contours $C_\pm$,
circling the upper- and lower-half planes, as shown in figure \ref{fig:wplane2}.
That is, omitting the $\rho,\rhobar$ arguments, we write
\be
 \c(\j,\Delta) = \int_0^1 \sigma d\sigma \left(
 \oint_{C} \frac{dw}{iw} \, \mu \,g\,N(\j,\Delta) \,F_{\j,\Delta}
 + \sum_{\pm} \oint_{C_\pm} \frac{dw}{i w} \,\mu\,g\,e^\pm_{\j,\Delta} \right)\,. \label{cj_inter_2}
\ee
Naturally, the extra functions $e^\pm_{\j,\Delta}$ should be eigenfunctions of the same Casimir equations.
These should also vanish at large $w$, in order to not re-introduce the problem there.
Since large $w$ corresponds to $0\ll \zbar\ll z\ll 1$, this singles out two of the eigenfunctions:
\be
 e^\pm_{\j,\Delta}(z,\zbar) =
  e^{\pm,1}_{\j,\Delta}\, \gpure_{\Delta+1-d,\j+d-1}(\zbar,z)
+e^{\pm,2}_{\j,\Delta}\, \gpure_{1-\Delta,\j+d-1}(\zbar,z)\qquad (w\to \infty)\,, \label{e_block}
\ee
where $e^{\pm,1}_{\j,\Delta}$ are unknown coefficients.
This ansatz vanishes like $w^{-\j}$ at large $w$ as can be seen from (\ref{gpure_text}).
Equation (\ref{cj_inter_2}), with the arcs at infinity dropped, thus holds for any choice of these coefficients.
We would like to find a choice such that the arcs at the origin also cancel, at least for sufficiently large spin,
that is
\be
 N(\j,\Delta) F_{\j,\Delta} -
 e^{+,1}_{\j,\Delta}\, \gpure_{\Delta+1-d,\j+d-1}
-e^{+,2}_{\j,\Delta}\, \gpure_{1-\Delta,\j+d-1} \propto w^{\j}  \qquad (w\to 0+i0)\,.
\ee
The minus signs account for the opposite orientation of the contours $C$ and $C_+$ near the origin.
Taking the coefficient of each 6 unwanted solution gives 6 constraints on two parameters.
To compute these constraints we performed the analytic continuation of the above
eigenfunctions to the origin $w\to 0$ using
a sequence of elementary monodromies: first, one lowers $w$ below $1/\sigma$, which takes $z$ counter-clockwise around 1 and changes the blocks $\gpure$
according to eq.~(\ref{cont_regge}). Then one takes $w$ below $1$ and interchange the arguments $z$ and $\zbar$ using eq.~(\ref{cont_spin}).
Finally, one takes $w$ below $\sigma$, which takes $\zbar$ counter-clockwise around 1 and is done using eq.~(\ref{cont_regge}) again.
Although the individual steps are simple, composing these three steps produces a complicated linear combination of all 8 solutions,
so we do not reproduce it here.  It needs to be added to the continuation of $F_{\j,\Delta}$
to $w<\sigma$, obtained using eq.~(\ref{cont_regge}) once.

Thus we obtain 6 constraints on two free parameters $e^{+,1/2}_{\j,\Delta}$.
This system is overconstrained so it is not obvious {\it a priori} that a solution exists.
It is not too surprising that the system is overconstrained, since a solution is at best expected when the spin is an integer.
It is still not obvious whether it should have any solution in that case, but the physical
analogy between the correlator and the $S$-matrix suggests that a solution should exist. Indeed, a solution does exist!
It is given as (\ref{e_block}) with
\be
 e^{+,1}_{\j,\Delta} = \tfrac14\kappa_{\j+\Delta} e^{i\pi(a+b+\frac{d-1}{2})}\,,\qquad
 e^{+,2}_{\j,\Delta}= e^{+,1}_{\j,\Delta}\times
 \frac{\Gamma\big(\tfrac{d}{2}-\Delta\big)\Gamma(\Delta-1)}{\Gamma\big(\Delta-\tfrac{d}{2}\big)\Gamma(d-\Delta-1)}\,.
\label{final_res}
\ee
and the complex conjugate expressions for $e^-$ below the axis.
Pleasantly, this is precisely the combination such that eq.~(\ref{e_block}) organizes into the block $G_{\Delta+1-d,\j+d-1}$, which is regular at $z=\zbar$
(but note the interchange of dimension and spin compared to the usual blocks.)
Also, most of the complicated factors from $N(\j,\Delta)$ have canceled out.
Furthermore, we find that, after adding $e^{-}$ below the axis and accounting for the phases in continuing the measure (\ref{measrho})
(from the unit circle, where it is real and positive),
the contributions from the four regions ($w\in (0,r)\cup (r,1)\cup(1,1/r)\cup(1/r,\infty)$), which all project onto the interval $0<z,\zbar<1$,
are all proportional to the same block!
Even more, the different regions are related to each other by the same phases as in the
positive-definite double discontinuity (\ref{dDisc_full}), times the absolute value of the measure,
exactly as we hoped for in section \ref{sec:review}!
Finally, the contribution from the cuts with $w<0$ produce a similar result but with an extra factor $(-1)^{\j}$.

It would be nice to understand more deeply why the above solution exists at all, perhaps by obtaining it through some
integral representation. 
In the rest of this paper we concentrate on its implications.

\subsection{Final result}

Our final result for the $s$-channel OPE coefficients, based on (\ref{final_res}), can thus be written as:
\be
 \c(\j,\Delta) = \c^t(\j,\Delta) + (-1)^{\j} \c^u(\j,\Delta) \label{fg_tu}
\ee
where each of the two channel contribution is an integral over a causal diamond:
\be
\framebox{\begin{minipage}[c][4ex][c]{.8\textwidth}\centering
$\c^t(\j,\Delta) =
\frac{\kappa_{\j+\Delta}}{4}\int_{0}^1 dz d\zbar\, \mu(z,\zbar)\, G_{\Delta+1-d,\j+d-1}(z,\zbar)\,\dDisc\big[\GG(z,\zbar)\big]\,.$
 \end{minipage}} \label{fg}
\ee
The $u$-channel contribution $\c^u$ is the same but with the integration ranging from $-\infty$ to 0
and the double discontinuity taken around $z=\infty$. Equivalently, it is the same but
with operators $1$ and $2$ interchanged.

Equation (\ref{fg}) is the main result of this paper.
It expresses the OPE coefficients in the $s$-channel ($x_1\to x_2$), isolated on the left-hand-side, in terms
of an integral over the correlator. In this sense it ``inverts'' the OPE.

Most importantly for us, and in contrast with the Euclidean formula (\ref{eucl_inversion}) which also ``inverts'' the OPE,
this CFT Froissart-Gribov formula is manifestly analytic in spin.
Just like the $S$-matrix Froissart-Gribov formula, the formula only works when ${\rm Re}\,\j$ is large enough
that the arcs at infinity vanish.  The bounds reviewed in section \ref{ssec:toy disp}
guarantee this for $\j>1$ in any unitary CFT.
The integrand is positive definite, due to positivity of the double discontinuity (\ref{dDisc_full}) and the absolute value sign in the measure (\ref{measure}).\footnote{
Technically, although always a positive-definite function,
for large enough external dimensions $\Delta_i$ the double discontinuity becomes a singular distribution near $\zbar=1$,
in which case the correct mathematical notion of ``positive-definite'' is less clear to us.
Integrals like (\ref{Ibeta}) seem to remain positive even in this case.}

Assuming analyticity in spin, it was shown in \cite{Costa:2012cb} how to extend the $s$-channel OPE
into the Lorentzian region by replacing the sum over spins by an integral;
the existence of the above formula apparently justifies this step.

It is not possible to immediately verify the formula (\ref{fg}) by simply inserting the $s$-channel OPE on the right-hand side
and verifying it termwise, since the double discontinuity vanishes for any individual $s$-channel block,
precisely as in the example in the introduction.
What makes the formula possible is that in any unitary theory the OPE coefficients for different spins are not independent of each other:
this is necessary for the OPE to resum to something sensible in Lorentzian signature.
In the next section we provide nontrivial checks confirming that the formula is correct;
explicit checks in the 2D Ising model are also given in appendix \ref{app:Ising}.

\section{Application to operators with large spin} \label{sec:spin}

The analyticity in spin makes the CFT Froissart-Gribov formula (\ref{fg}) ideal to study
large spin operators and understand how they organize into analytic families.
Operators with large spin have been studied extensively in the literature as noted in introduction,
and here we will rederive and extend some of these results
using the above formula; in particular, instead of asymptotic expansions in $1/\j$, we find
sums which explicitly converge down to $\j=2$.

\subsection{Generating function}

The spectrum of the theory is encoded in the poles of the inversion
integral (\ref{fg}). These originate from the $z\to 0$ integration boundary.
In practice, it is really only the one-dimensional $\zbar$ integration which produces the OPE coefficients,
and it will be useful to make this manifest by integrating out the $\zbar$ variable
to define a generating function $C^\pm(z,\beta)$.

To avoid double-counting,
it is better to restrict the integration range in eq.~(\ref{fg}) to $\zbar>z$, at the cost of a factor of two. 
A further simplification is to decompose the block
into two pure power solutions $\gpure$ using eq.~(\ref{G_from_gpure}): the second solution ensures the symmetry (\ref{shadow})
between operators and their shadow, but we do not expect it
to contribute to the poles at $\Delta>d/2$ in general.  Thus, as far as the Euclidean
OPE data is concerned, we can rewrite the inversion formula as:
\be
c^t(\j,\Delta)\Big|_{\rm poles}=
\int_0^1 \frac{dz}{2z} z^{\frac{\j-\Delta}{2}}\left(
\int_z^1 d\zbar\,\kappa_{\j+\Delta}\,\frac{\mu(z,\zbar)\, \gpure_{\Delta+1-d,\j+d-1}(z,\zbar)}{z^{\frac{\j-\Delta}{2}-1}}
\,\dDisc\big[\GG(z,\zbar)\big]\right). \label{C_gen}
\ee
The $z$ integral is now trivial, since it simply converts 
a term $z^{\frac{\tau}{2}}$ in the expansion of the parenthesis into a pole
$1/(\tau+\j-\Delta)$, that is, to the exchange of an operator of twist $\Delta-\j=\tau$.
Thus the $z$ dependence as $z\to 0$ simply tracks the twist of operators
and we can regard the parenthesis as a generating function.

The parenthesis depends on three variables: $z,\j,\Delta$. If we focus on the $z\to 0$ limit,
however, the nontrivial dependence is only through $z$ and the sum $\beta=\Delta+\j$.
This combination is called the ``conformal spin'' in the literature because it corresponds
to the Casimir invariant of the SL${}_2$(R) conformal symmetries of a null ray.
Performing the integral in that limit we thus define a SL${}_2$(R) generating function:
\be\label{C_coll}
\framebox{\begin{minipage}[c][4ex][c]{.6\textwidth}\centering{$
 C^{t}(z,\beta) \equiv \int_z^1 \frac{d\zbar(1-\zbar)^{a+b}}{\zbar^2} \,\kappa_\beta\,k_\beta(\zbar) \,\dDisc\big[\GG(z,\zbar)\big].$}\end{minipage}}
\ee
Physically, this integral projects onto SL${}_2$(R) primaries with respect to the null direction $\zbar$,
but it does not subtract descendants along the $z$ direction.
On the other hand
these can be automatically subtracted by expanding the block in eq.~(\ref{C_gen}) to subleading powers 
in $z$.  In fact, according to eq.~(\ref{twist_expansion}), the expansion contains only SL${}_2$(R) blocks with shifted value of $\beta$,
hence the OPE data for conformal primaries can be expressed entirely in terms of the SL${}_2$(R) generating function:\footnote{The coefficient $B^{(m,k)}_{\j,\Delta}$ have poles
at twists $\Delta-\j-d=0,1,2,\ldots$ as predicted by eq.~(\ref{C_residues}). These are removed when taking the combination
in that equation.}
\be
c^t(\j,\Delta)\Big|_{\rm poles} = \int_0^1 \frac{dz}{2z} z^{\frac{\j-\Delta}{2}}
\left(\sum_{m=0}^\infty z^{m}  \sum_{k=-m}^m B^{(m,k)}_{\j,\Delta} C^t(z,\j{+}\Delta{+}2k)\right), \label{C_series}
\ee
where the $B^{(m,k)}_{\j,\Delta}$ are rational functions of $\j,\Delta$ that are computable
recursively using the quadratic Casimir equation.  
Only finitely many terms in $m$ are needed to extract any given OPE coefficient: just enough to subtract all the descendants from lower twist primaries.
We note also that the process of extracting the coefficient of a given power of $z$ in eq.~(\ref{C_coll}) commutes with
doing the $\zbar$ integral (for $\j>1$), whose boundary can then be taken to be $0$.\footnote{
This follows from the fact that the correlator $\dDisc\big[\GG(z,\zbar)\big]$ is bounded as $z=\zbar\to 0$,
so the lower-endpoint of the range of integration can only produce powers of $z$ higher than $z^{(\Delta+\j)/2-1}$,
independently of $m$ and $k$ in eq.~(\ref{C_series}). This is never of the form $z^{(\Delta-\j)/2}$ for any $\j>1$.}

Finally, let us record the precise relation between the generating function (\ref{C_coll}) and OPE coefficients.
According to the inversion formula (\ref{fg_tu}),
operators with even and odd spins constitute two independent analytic families,
and the $t$ and $u$ channel contributions should be added with a relative sign depending on whether
the desired operator has even or odd spin,
\be
 C^{(\pm)}(z,\beta) = \sum_m \,C^{(\pm)}_m(\beta) \,z^{\frac12\tau^{(\pm)}_m(\beta)}\,,
 \qquad C^{(\pm)} \equiv C^t\pm C^u\,.  \label{generating_fct}
\ee
The expansion coefficients $C^{(\pm)}_m(\beta)$ and $\tau^{(\pm)}_m(\beta)$ then encode
the OPE coefficients and twists of the operators.
Both are analytic in $\beta$ and to obtain physical operators one should restrict to those values of $\beta$
for which the spin $\frac12(\beta-\tau)$ happens to be an integer. 
Actually, $C$ is not quite the OPE coefficient, because, according to eq.~(\ref{C_residues}), the $\Delta$-residue of $1/(\Delta-\j-\tau(\beta))$ should be taken with $\j$ fixed. This produces a extra Jacobian factor:
\be
 f_{12\OO}f_{43\OO} =
 \left(1-\frac{d\tau^{(\pm)}_m(\beta)}{d\beta}\right)^{-1} C^{(\pm)}_m(\beta) \Bigg|_{\frac{\beta-\tau^{(\pm)}_m(\beta)}{2}=\j}
\ee
where the sign $(\pm)$ depends on the parity of $\j$.  Precisely this formula, and the existence of a function $C(z,\beta)$,
has been deduced empirically before from large-$\j$ expansions, see \cite{Alday:2015eya,Simmons-Duffin:2016wlq}.
For operators of subleading twist one should replace $C$ by $c$ to account for the subtraction of descendants according to eq.~(\ref{C_series}).

\subsection{Vaccum exchange (or exchange of pure power-law)}

A simple but important case where the integral (\ref{C_coll}) can be done analytically is the unit operator in the $t$-channel.
One then expects $s$-channel OPE coefficients to be products of $\Gamma$-functions.
In fact, since one can vary the dimensions of the external operators, this gives an infinite set of integrals which can be done analytically.
Taking the pair of operators (4,3) to be the same as (1,2), the vacuum contribution to the generating function,
is, according to eqs.~(\ref{t-channel_OPE}) and (\ref{C_coll}),
\be
 C^+(z,\beta)\supset z^{\frac{\Delta_1+\Delta_2}{2}} I_{-\Delta_1-\Delta_2}^{(a,a)}(\beta) \qquad \mbox{(vacuum exchange)}\,,
 \label{unit_contribution}
\ee
where as before $a=\frac{\Delta_2-\Delta_1}{2}$,
and using a standard Euler-type representation for the hypergeometric function $k_\beta$, we could evaluate the following general
integral:
\be\label{Ibeta}\begin{aligned}
I_{\tau'}^{(a,b)}(\beta)
&\equiv \int_0^1 \frac{d\zbar}{\zbar^2} (1-\zbar)^{a+b}\kappa_\beta k_\beta(\zbar)\, \dDisc\!\left[
\left(\frac{1-\zbar}{\zbar}\right)^{\frac{\tau'}{2}-b} (\zbar)^{-b}\right]
\\ &= \frac{1}{\Gamma\big(-\tfrac{\tau'}{2}-a\big)\Gamma\big(-\tfrac{\tau'}{2}+b\big)}\times
\frac{\Gamma\big(\tfrac{\beta}{2}-a\big)\Gamma\big(\tfrac{\beta}{2}+b\big)}{\Gamma\big(\beta-1\big)}
\times \frac{\Gamma\big(\tfrac{\beta}{2}-\tfrac{\tau'}{2}-1\big)}{\Gamma\big(\tfrac{\beta}{2}+\tfrac{\tau'}{2}+1\big)}\,.
\end{aligned}\ee
This formula will be used extensively in this section.

Some comments are in order.
First, the integral vanishes when the exponent $\tau'$ differs from $2b$ or $-2a$ by a nonnegative integer,
in agreement with the discussion in subsection \ref{ssec:largeN}: physically these exponents
represent ``double trace'' $t$-channel operators of twist $\Delta_2+\Delta_3$ or $\Delta_1+\Delta_4$ plus even integers.
Note also that one might be concerned with convergence near $1$ for sufficiently negative exponent $\tau'$, but the double discontinuity operation actually ensures that the integral converge near $\zbar=1$.
This can be understood from the derivation in the preceding section, where the double-discontinuity arose from the contour integrals in fig.~\ref{fig:wplane2}, which smoothly avoid the branch point $\zbar=1$.
This ensures that the integral is an analytic function for all $\tau'$, which the
explicit formula indeed makes manifests (except for poles in the last factor, which are due to divergences around 0).
The preceding formula is therefore correct even for negative $\tau'$, even though the integral would seem to diverge near 1.

The nonzero result for negative integer exponent, where the double-discontinuity naively vanishes,
can be interpreted as a $\delta$-function terms at $\zbar=1$:
it is a familiar result that the discontinuity of a propagator $1/x_{23}^2$ is a $\delta$-function.
In the same sense, it is not too surprising to learn from the above formula that
that $1/(1-\zbar)\sim 1/(x_{23}^2x_{14}^2)$ has a double-discontinuity, because it contains two propagators.
In practice, to deal with singular correlators, one can simply use eq.~(\ref{Ibeta}) to integrate the singular part analytically,
leaving an unambiguous remainder.

Incidentally, a converse to formula (\ref{Ibeta}) appeared recently in \cite{Simmons-Duffin:2016wlq} while this work
was being completed, where the following sum of SL${}_2$(R) blocks was derived:
\be
 \sum_{m=0}^\infty I^{(a,b)}_{\tau'}(\beta_0+2m) k_{\beta_0+2m}(\zbar) = \left(\frac{1-\zbar}{\zbar}\right)^{\frac{\tau'}{2}-b} (\zbar)^{-b}
 - \mathcal{R}^{(a,b)}_{\tau',\beta_0}(\zbar), \label{sl2_sum}
\ee
where $\mathcal{R}$ is an explicitly known remainder, with the property that is has
a vanishing double-discontinuity around $\zbar=1$.\footnote{
For identical operators, we record for reference the expression from \cite{Simmons-Duffin:2016wlq}:
\be
 \mathcal{R}_{\gamma,\beta_0}^{(0,0)}(1-z) =
 \frac{1}{\Gamma\big(-\tfrac{\gamma}{2}\big)^2}
 \frac{\Gamma\big(\tfrac{\beta_0-\gamma}{2}-1\big)}{\Gamma\big(\tfrac{\beta_0+\gamma}{2}\big)}
\sum_{k=0}^\infty
 \frac{\partial}{\partial k} \frac{\Gamma\big(\tfrac{\beta_0}{2} +k\big)}{(\tfrac{\gamma}{2}-k)\Gamma(k+1)^2
 \Gamma\big(\tfrac{\beta_0}{2} -k-1\big)} \left(\frac{z}{1-z}\right)^k\,. \label{remainder_expr}
\ee}
This property was termed ``Casimir-regular'' in \cite{Simmons-Duffin:2016wlq}.
This is a nice consistency check which confirms that the SL${}_2$(R) inversion integral (\ref{C_coll}) indeed ``inverts'' the coefficient of
$k_\beta$.

At large spin or equivalently large $\beta$,
the collinear block $k_{\beta}(\zbar)$ behave like $(\rhobar)^{\beta/2}$ and so the inversion
integral is dominated by $\zbar\to 1$. There the $t$-channel blocks vanish with an exponent governed by their twist.
One thus expect the large-$\beta$ limit of the $s$-channel OPE coefficients to decay with the twist $\tau'$
of $t$-channel exchanged operators,
and, indeed, the above integral decays like $1/\beta^{(\tau'+a-b+3/2)}$.

It follows that, at large $\beta$, the $t$-channel unit contribution to the generating functional (\ref{unit_contribution})
can't be cancelled by any other operator, if the theory has a twist gap above the unit operator (which is always the case for a unitary theory in $d>2$).
This is precisely the conclusion reached in \cite{Komargodski:2012ek,Fitzpatrick:2012yx}:
operators with twist arbitrarily close to $\Delta_1+\Delta_2$ must exist at large spin.
The same argument holds for subleading powers of $z$, leading
to primaries of dimension $\Delta_1+\Delta_2+2m+\j$. 
Following \cite{Simmons-Duffin:2016wlq}, we refer to these as double-twist operators $[12]_m(\j)$.
A central focus of the analytic bootstrap program is to understand the $1/\beta$ corrections to these multi-twist families,
to which we now turn.

\subsection{Systematics of large-$\j$ corrections}

The $t$-channel unit contribution (\ref{unit_contribution})
gets corrected, at subleading orders in $1/\beta$, by a convergent sum over $t$-channel primaries of subleading twist,
each decaying like $1/\beta^{\tau'}$.
First we focus on an individual primary, and then we discuss the effects of the infinite summation.

Our starting point will be the generating functional (\ref{C_coll}) with the correlator represented by
its $t$-channel OPE. Accounting for the prefactor in the crossing relation (\ref{t-channel_OPE}),
\be\begin{aligned}
 C(z,\beta) &= \sum_{\j',\Delta'}f_{14\OO'}f_{23\OO'}
\int_z^1 \frac{d\zbar(1-\zbar)^{a+b}}{\zbar^2} \kappa_{\beta}k_{\beta}(\zbar) 
\,\dDisc\!\!\left[
\frac{(z\zbar)^{\frac{\Delta_1+\Delta_2}{2}}G_{\j',\Delta'}(1-\zbar,1-z)}{\big[(1-z)(1-\zbar)\big]^{\frac{\Delta_2+\Delta_3}{2}}}\right]. \label{t-channel-C}
\end{aligned}\ee
The sum converges (at fixed $z$) since all sampled values of $\zbar$ lie within the convergence radius of the OPE.
We now first tentatively take the $z\to 0$ limit term by term---this will not be completely correct, but almost!

As discussed in appendix \ref{app:cross}, for each block there are two towers of terms in the $z\to 0$ limit,
starting from two exponents
$\tau_1=\Delta_1+\Delta_2$ and $\tau_2=\Delta_3+\Delta_4$:
\be \label{cross_expansion_text}
z^{\frac{\Delta_1+\Delta_2}{2}}
G_{\j',\Delta'}(1{-}\zbar,1{-}z)=
 \sum_{i=1,2}\,
 \sum_{n=0}^\infty z^{\frac12\tau_i+n}\times H_{\j',\Delta'}^{(i),n}(1-\zbar)\,.
\ee
The computation of the functions $H$ in general dimension is detailed in appendix.
There is a slight issue when the initial and final operator pairs are identical, as is needed to get matrix elements between identical
double-twist operators: then the two exponents $\tau_i$ are identical and logarithms of $z$ appear.
The $z\to 0$ expansion  must then be rewritten slightly:
\be \label{cross_expansion_log_text}\begin{aligned}
z^{\frac{\Delta_1+\Delta_2}{2}}
G_{\j',\Delta'}(1{-}\zbar,1{-}z)
= \sum_{n=0}^\infty z^{\frac{\Delta_1+\Delta_2+2n}{2}}
\times \left(\tfrac12\log z\,H_{\j',\Delta'}^{\slog,n}(1-\zbar)+H_{\j',\Delta'}^{\sreg,n}(1-\zbar)\right).
\end{aligned}\ee
The $\frac12\log z$ term is interpreted, to first order in the anomalous dimension,
as a shift to the dimension of the double-twist operators. At large spin, one only needs the limit $\zbar\to 1$ of the $H$ functions, which
can be determined simply from SL${}_2$(R) blocks (see eq.~(\ref{hypers_at_one})) and reads
\be
H_{\j',\Delta'}^{({\rm log}),0}(1-\zbar) \approx -2\frac{\Gamma(\Delta'+\j')}{\Gamma\big(\tfrac{\Delta'+\j'}{2}\big)^2}(1-\zbar)^{\frac{\Delta'-\j'}{2}}(1+O(1-\zbar))\,.
\ee
Plugging into (\ref{Ibeta}) and dividing by the unit operator contribution in eq.~(\ref{unit_contribution}), we thus obtain the first-order
correction to double-twist anomalous dimensions from  exchange of a single $t$-channel operator of spin $\j'$ and dimension
$\Delta'$:
\be\begin{aligned}
\Delta_{[12]_0}(\j) -(\Delta_1+\Delta_2 +\j)  &\approx -2f_{11\OO'}f_{22\OO'}\frac{\Gamma(\Delta'+\j')}{\Gamma\big(\tfrac{\Delta'+\j'}{2}\big)^2}
\frac{I_{\Delta'-\j'-\Delta_1-\Delta_2}^{(a,a)}(\beta)}{I_{-\Delta_1-\Delta_2}^{(a,a)}(\beta)}
 \\ &\approx -2f_{11\OO'}f_{22\OO'}\frac{\Gamma(\Delta'+\j')\Gamma(\Delta_1)\Gamma(\Delta_2)}{\Gamma\big(\tfrac{\Delta'+\j'}{2}\big)^2
\Gamma\big(\Delta_1-\frac{\Delta'-\j'}{2}\big)\Gamma\big(\Delta_2-\frac{\Delta'-\j'}{2}\big)}\frac{1}{\big(\beta/2\big)^{\Delta'-\j'}}
\label{leading_term_1/j}
\end{aligned}\ee
where $\beta=\Delta_{[12]_0}(\j)+\j$ is the conformal spin, and the approximation signs are up to subleading orders in $1/\beta$
(coming both from the truncation of the $(1-\zbar)$ series and of the integral (\ref{Ibeta})).

Eq.~(\ref{leading_term_1/j}) is in perfect agreement with formulas from sections 3 and 4 of \cite{Komargodski:2012ek}.
This is a nice check of all the factors in our inversion formula (\ref{fg}).

A comment is in order regarding the $u$-channel cut contribution, neglected in the above,
but which is to be added with an extra sign $(-1)^\j$ in eq.~(\ref{fg_tu}).
The $u$-channel term would be absent in certain situations where
the operators are distinct, but in general
one should sum up the $t$- and $u$-channel contributions prior to dividing by the vacuum exchange.
Thus even and odd spins  generally describe two totally independent analytic families of operators.
In the case where the operators $1,2$ are identical, the above formula remains valid  but should be restricted to even spins.

In the case of stress tensor exchange, the OPE coefficient is fixed by symmetries.
With our blocks normalized as in (\ref{normalization}) and the $TT$ normalization $C_T$ defined as in \cite{ElShowk:2012ht},
the OPE coefficients reads $f_{iiT}= \frac{\Delta_i d}{2\sqrt{C_T}(d-1)}$. Plugging it into (\ref{leading_term_1/j}) it then
also agrees with \cite{Komargodski:2012ek,Fitzpatrick:2012yx}.

The main difference so far compared with these works,
is that the inversion formula applies to each individual operator in the even/odd spin families, as opposed to giving
only their average large-spin properties.
In particular, this establishes the existence of each individual double-twist operator (for sufficiently large spin that we can ignore the possibility of coefficients summing up to zero or very large anomalous dimensions), and it explains why they organize into analytic families in the first place.

\subsubsection{Subleading powers: individual block}

To generate subleading terms in $1/\j$ from a given block is now straightforward: one simply expands the function
$H_{\j',\Delta'}^{\slog,0}(1-\zbar)$ to higher orders in
$(1-\zbar)/\zbar$ and use the analytic integral (\ref{Ibeta}).
This can be done to any desired order using the quartic equation satisfied by $H$, see appendix \ref{app:cross}.
This however only produces an asymptotic expansion in $1/\j$, because the series in $(1-\zbar)/\zbar$ does not converge for $\zbar<\frac12$.  The inversion formula makes it possible to do better, since in principle one can just do the $\zbar$ integral numerically and conceptually there is no need to expand in $1/\j$.

To illustrate this in a simple concrete example, consider the exchange of a $t$-channel primary of spin $\j'=0$.
The $z\to 0$ limit of the $t$-channel block, $H_{\j',\Delta'}^{({\rm log}),0}(1-\zbar)$, can be evaluated analytically
in terms of a hypergeometric function given in eq.~(\ref{exact_j0}).
Thus the contribution of an individual $\j'=0$ primary to the double-twist anomalous dimension is,
restricting eq.~(\ref{t-channel-C}) to lowest twist:
\be\begin{aligned}
\Delta_{[12]_0}(\j) -(\Delta_1+\Delta_2 +\j)
&\approx
-2f_{11\OO'}f_{22\OO'}\frac{\Gamma(\Delta')}{\Gamma\big(\tfrac{\Delta'}{2}\big)^2I_{-\Delta_1-\Delta_2}^{(a,a)}(\beta)}
\int_0^1 \frac{d\zbar\,(1-\zbar)^{\Delta_2-\Delta_1}}{\zbar^2}
\kappa_{\beta}k_{\beta}(\zbar)
\\ & \hspace{20mm}\times
\,\dDisc\!\left[\frac{{}_2F_1\big(\tfrac{\Delta'}{2},\tfrac{\Delta'}{2},\Delta'-\tfrac{d-2}{2},1-\zbar\big)}{(1-\zbar)^{\Delta_2-\Delta'/2}}
\zbar^{\frac{\Delta_1+\Delta_2}{2}}\right].
\end{aligned}\ee
The approximation sign is only because we have used the coefficient of $\tfrac12\log z$ in the generating function as a proxy for the anomalous dimension, as above, but this formula exactly gives the coefficient of $\tfrac12\log z$.
(This proxy will be relaxed shortly.)

Expanding the integral at large spin $\beta$ we reproduce, for example, the first few terms of the large spin expansion
given for four scalars in the 3D Ising model in eq.~(5.6) and footnote 14 of \cite{Alday:2015ota} (with $j^2_{\rm there}=\frac14\beta(\beta-2)_{\rm here}$). Thus the above integral indeed resums the $1/\j$ expansion to all orders! It converges for $\beta> d-\Delta_1-\Delta_2$, which includes all operators of spin $\j\geq 2$
due to unitarity bounds.

\subsubsection{Beyond leading-log: exact sum rule and application to 3D Ising}

If one could sum up all $t$-channel primaries, the convergent expansion (\ref{t-channel-C}) would reproduce exactly
all $s$-channel OPE coefficients.  However, in practice, one must
address the fact that the infinite sum over primaries does not commute with taking
the $z\to 0$ limit of each term.
This is because the true $z\to 0$ limit of the generating function involves power-laws $z^{\tau/2}$,
in contrast with individual $t$-channel blocks, which have at most single logarithms times double-twist powers.

A simple solution is to subtract a known sum, such as the SL${}_2$(R) sum in eq.~(\ref{sl2_sum}).
Let us restrict, for notational simplicity, to the case where all four operators are the same scalar
$\sigma$, and to the lowest twist; generalization will be straightforward.
From the $t$-channel sum (\ref{t-channel-C}) we know that
\be
 C_0(\beta) z^{\Delta_\sigma+\frac12\gamma_0(\beta)}+\mbox{subleading in $z$}
 = \sum_{\j',\Delta'} f_{\sigma\sigma\OO'}^2\tilde{I}_{\j',\Delta'}(z,\beta), \label{Csum1}
\ee
where $C_0(\beta)$ and $\gamma_0(\beta)$ are the lowest-twist OPE coefficicent and anomalous dimension
appearing in eq.~(\ref{generating_fct}),
and $\tilde{I}$ stands for the universal (theory-independent) integral of a block in eq.~(\ref{t-channel-C}).
Subtracting eq.~(\ref{sl2_sum}) with $\zbar\to 1{-}z$ (and an arbitrary $\beta_0$) and dividing by $z^{\Delta_\sigma}$,
this replaces the exponent on the left by  a constant and logarithm:
\be
A(\beta)+B(\beta)\tfrac12\log z
=
\sum_{\j'=0}^\infty \Bigg[
-C_0(\beta)\, I^{(0,0)}_{\gamma_0(\beta)}\!(\beta_0{+}2\j')\,k_{\beta_0{+}2\j'}(1{-}z)
+
\frac{1}{z^{\Delta_\sigma}}\sum_{\Delta'}f_{\sigma\sigma\OO'}^2\tilde{I}_{\j',\Delta'}(z,\beta)
\Bigg]_{z\to 0}.
\label{Csum2}
\ee
The point is that we can now take the $z\to 0$ limit termwise and obtain two equations
giving the coefficients $A,B$ as \emph{convergent} sums over $t$-channel primaries.
The coefficients $A$ and $B$ are then directly related to the $s$-channel OPE coefficients and anomalous dimension
through the expression for the remainder $\mathcal{R}$ given in eq.~(\ref{remainder_expr}):
\be\begin{aligned}
 A(\beta)+B(\beta)\tfrac12\log z &\equiv C_0(\beta)\mathcal{R}_{\gamma_0(\beta),\beta_0}^{(0,0)}(1-z)\Big|_{k=0}
= C_0(\beta)\left(1+\tfrac{\gamma_0(\beta)}{2}\log z+ O(\gamma_0^2)\right) \,. \label{sum_rule}
\end{aligned}\ee
Thus, when the anomalous dimension is small, the sum rule in eq.~(\ref{Csum2}) reduces to the naive procedure of
extracting OPE coefficients and anomalous dimensions from regular and logarithmic terms, as was done above,
but in general the $k=0$ term in $\mathcal{R}$ incorporates corrections of order $\gamma_0^2$.
Note also that even though $C$ and $\gamma_0$ appear on both sides of the equation, their effect is much more important
on the left-hand side, so the equation can be solved iteratively.

At subleading twists, the same logic gives rise to two sum rules for each exponent $\tau=2\Delta_\sigma+2m$ where $m=0,1,2,\ldots$.
The number of subtraction needed for convergence is equal to the number of $s$-channel
operators with twist less than $2\Delta_\sigma+2m$ at that particular value of $\beta$, which we generally expect to be finite.

Somewhat reminiscent sum rules have been used recently in Mellin space \cite{Gopakumar:2016wkt,Gopakumar:2016cpb}.
These authors expand the correlator in terms of Witten diagrams instead of conformal blocks and then require
that spurious ``double trace'' operators of twist $2\Delta+2m$ cancel out. It would be interesting to better understand the connection.

Using the numerical data for the 3D Ising model provided graciously in \cite{Simmons-Duffin:2016wlq},
we could numerically check the above sum rule (\ref{sum_rule}); including the operators
from the $[\sigma\sigma]_0$, $[\sigma\sigma]_1$ and $[\epsilon\epsilon_0]$ families tabulated in the appendices of that paper,
we checked the sum rule for the $s$-channel stress tensor to $10^{-3}$ accuracy for its dimension and OPE coefficient.
This is impressively accurate although not significantly different than the quality of the asymptotic expansions already considered in \cite{Simmons-Duffin:2016wlq}. We leave it as an open question to identify which operators must be included
to increase the precision beyond this point.

An interesting possibility is to use the convergent sums to control the errors.
For example, schematically, an alternative way to extract the lowest twist at a given $\beta$ is from
\be
 -\gamma(\beta)= \lim_{z\to 0} \frac{(2\Delta_\sigma-2z\partial_z) C(z,\beta)}{C(z,\beta)}\,.
\ee
At finite $z$, both the numerator and denominator are convergent $t$-channel sums,
and evaluating the sum at finite but small $z$ (say $10^{-3}$) the error will be proportional to $z$.
Both the numerator and denominator are generically sums of positive terms.
Since the denominator starts with 1, the effect of any given $t$-channel primary
is stronger on the numerator than denominator. For the right-hand-side to not exceed
$2\Delta_\sigma-1\approx 0.036$ for the stress tensor ($\beta=5$),  which the $\epsilon$-exchange already nearly comes close
to saturating, then gives an upper bound on the remaining operator contributions.
When we change the values of $\beta$, these get smaller,
with the higher-dimension contributions decaying more rapidly with $\beta$, thus allowing to bound the uncertainties on other operators using the error on the stress tensor.

The main novelty of the inversion formula (\ref{fg}), compared with formulas from
refs.~\cite{Alday:2015ota} and \cite{Simmons-Duffin:2016wlq} which include similar physics,
is that conceptually it produces convergent sums that are valid for any individual spin $\j>1$,
as opposed to inverting a crossing equation as a series in $1/\j$. Although this does not seem to make a big numerical
difference for the 3D Ising model, this does explain conceptually why the $1/\j$ expansions
of \cite{Alday:2015ota,Simmons-Duffin:2016wlq} appear to work all the way down to the stress tensor.
It will be interesting to see how this helps make error estimates
that can be used in practice by the numerical bootstrap program.

\subsubsection{Quadruple cut equation: Large spin in both channels}

We conclude this discussion with a brief analysis of the interplay between operators of large spin in both channels.
Just like the double discontinuity around $\zbar=1$, which kills
individual $s$-channel blocks and allows to focus on the analytic-in-spin part, by taking a further double-discontinuity at $z=0$
one can focus on the part which is analytic-in-spin in both channels.

We do this by defining a generating function $C(\beta',\beta)$ that depends on two conformal spins.
Specifically, we integrate $C(z,\beta)$ over $z$ using a measure similar to $\zbar$ in eq.~(\ref{C_coll}).
Restricting, for simplicity, to identical external operators of dimension $\Delta$, we thus define:
\be
 C(\beta',\beta)
\equiv \int_0^1 \frac{dz}{(1-z)^{2-\Delta}} \kappa_{\beta'}k_{\beta'}(1-z)
\int_z^1 \frac{d\zbar}{\zbar^2}  \kappa_{\beta}k_\beta(\zbar)\,\,{\rm qDisc}\!\left[\frac{1}{z^\Delta}\GG(z,\zbar)\right].
\label{Cbb}
\ee
Here ${\rm qDisc}$ is the \emph{quadruple} discontinuity: the
double-discontinuity around $\zbar=1$ followed by the double-discontinuity around $z=0$.
At large spins $\beta'$, $\beta$, the integral is dominated by the corner $(z,\zbar)\to (0,1)$,
which is the usual double-lightlike limit.

Plugging in the expansion (\ref{generating_fct}), this can be interpreted as a sum over families
of operators in the $s$-channel:
\be
 C(\beta',\beta) = \sum_{m} C_m(\beta)\left(I^{(0,0)}_{\gamma_m(\beta)}(\beta')+\mbox{subleading}\right), \label{Cbb1}
\ee
where $\gamma_m=\tau_m-2\Delta$ are anomalous dimension (defined relative to double-twist operators)
and the (known) omitted terms originate only from the difference between $z^{\gamma/2}$ and $(z/(1-z))^{\gamma/2}$
in the small $z$ expansion and are subleading at large $\beta'$.
On the other hand, the factors to the power $\Delta$ in eq.~(\ref{Cbb}) have been chosen so that
the formula is crossing symmetrical under interchange of $\beta$ and $\beta'$,
and so it can also be interpreted as a sum over $t$-channel operators:
\be
C(\beta',\beta) = \sum_{m} C'_{m}(\beta')\left(I^{(0,0)}_{\gamma'_{m}(\beta')}(\beta)+\mbox{subleading}\right). \label{Cbb2}
\ee
What can be learnt from equating (\ref{Cbb1}) and (\ref{Cbb2})?

As an example, in the 3D Ising model, considering the $\sigma\sigma\sigma\sigma$ correlator
and taking $\beta'\gg \beta\gg 1$ to project onto the lowest  $s$-channel trajectory $[\sigma\sigma]_0$,
the equality reduces to:
\be
\frac{\tilde{C}_0(\beta)}{\Gamma(-\tfrac12\gamma_0(\beta))^2}\frac{1}{(\beta'/2)^{\gamma_0(\beta)}}
\times\Big(1+\mbox{subleading}\Big)
=
\sum_{m} \frac{\tilde{C}'_m(\beta')}{\Gamma(-\tfrac12\gamma_m'(\beta'))^2}\frac{1}{(\beta/2)^{\gamma'_m(\beta')}},
\ee
where $\tilde{C}_0(\beta)\equiv \frac{2^\beta}{(\beta/2)^{\frac32}}C_0(\beta)$, and again we omit computable $1/\beta$ corrections.  The double-twist anomalous dimension $[\sigma\sigma]_0$ vanishes at large $\beta$ (see eq.~(\ref{leading_term_1/j})):
$\gamma_0(\beta)\propto -f_{\sigma\sigma\epsilon}^2/\beta^{\Delta_\epsilon}$.
The left-hand side can thus be expanded into powers $1/\beta^{(m\Delta_\epsilon)}$ where $m=2,3,\ldots$
Comparing with the right-hand-side, where the power of $\beta$ correspond to the twist of operators,
one concludes that multi-twist families $\{[\epsilon\epsilon],[\epsilon\epsilon\epsilon],\ldots\}$ of twist $m\Delta_{\epsilon}$
must exist, and one also predicts their (averaged) OPE coefficients.
For example, from the $m=2$ case,
the $[\epsilon\epsilon]$ OPE coefficient must approach a constant asymptotically.  Physically, this
ensures that the $t$-channel OPE reproduces the correct term $\tfrac18\gamma^2\log^2z$ predicted by exponentiation
of the leading anomalous dimension (this was also discussed recently in \cite{Simmons-Duffin:2016wlq}). 
For the $[\epsilon\epsilon\epsilon]$ family, the formula predicts OPE coefficients which grow like $\log(\beta')$, etc.  

Finally, let us mention another interesting situation where the quadruple discontinuity seems particularly
apt at capturing the physics: the interplay between isolated lowest-twist trajectories in both channels,
which gives, in a general CFT:
\be
\frac{\tilde{C}_0(\beta)}{\Gamma(-\tfrac12\gamma_0(\beta))^2}\frac{1}{(\beta'/2)^{\gamma_0(\beta)}}
=
\frac{\tilde{C}'_0(\beta')}{\Gamma(-\tfrac12\gamma_0'(\beta'))^2}\frac{1}{(\beta/2)^{\gamma'_0(\beta')}}.
\ee
Taking the logarithm on both sides, one sees that the right-hand-side can grow at most linearly with $\log(\beta)$.
Imposing this on the left-hand-side, one concludes that, for such isolated trajectories, there must exist constants such that:
\be\begin{aligned}
 \lim_{\beta\to \infty} \gamma_0(\beta) &= 2\Gamma_{\rm cusp}\log(\beta/2)+\gamma_\infty  +\mbox{power suppressed},
\\
 \lim_{\beta'\to \infty} \frac{\tilde{C}'_0(\beta')}{\Gamma(-\tfrac12\gamma_0'(\beta'))^2}
  &= c\times (\beta'/2)^{-\gamma_\infty} + \mbox{power suppressed}.
\end{aligned}\ee
This gives a simple proof that the well-known logarithmic scaling behavior of gauge theories is the most general possibility
consistent with crossing symmetry, as originally proved in \cite{Alday:2013cwa}.
The limit for the OPE coefficient, when $\Gamma_{\rm cusp}\neq 0$, also agrees with the result there.
(The extension remains relatively simple when there is only one operator in one channel but many in the other,
see \cite{Alday:2016mxe}.)  
It would be nice to understand how subleading twist trajectories in both channels interact with each other,
perhaps combining the quadruple discontinuity with the
methods of \cite{Alday:2016jfr,Alday:2016njk}; we leave this for the future.

\section{Application to AdS bulk locality and Witten diagrams}\label{sec:gravity}

The inversion formula (\ref{fg}), involving a bounded and positive definite integrand
dominated by single-trace operators, seems an ideal tool to analyze CFTs with gravity duals.
As an application we derive here upper bounds on higher-derivative interactions.

\subsection{Bounding heavy operator contributions as a function of spin}

Consider a theory with a sparse spectrum of single-trace operators, characterized by a large gap $\Delta_{\rm gap}$,
which for simplicity we define here as the lowest twist of single-trace primaries with spin $\j>2$.
The contributions to the $t$-channel OPE can be separated into light and heavy operators according to their twist.
Double-trace primaries contribute to the double discontinuity only at subleading order in $1/N$,
as discussed in section \ref{ssec:largeN},
and the heavy contribution to the double discontinuity can be estimated as in eq.~(\ref{heavy_contribution}):
\ba
 \dDisc \GG\big|_{\rm heavy} &=&
 \mbox{(prefactor)}\sum_{\Delta-\j> \Delta_{\rm gap}} c_{\j',\Delta'}
 \left(\frac{1-\sqrt{\rho}}{1+\sqrt{\rho}}\right)^{\Delta' +\j'}
 \left(\frac{1-\sqrt{\rhobar}}{1+\sqrt{\rhobar}}\right)^{\Delta' -\j'}\times \mbox{phases}
\nonumber\\
&\leq& e^{-\Delta_{\rm gap}(\sqrt{z}+\sqrt{\zbar})}\,. \label{upper_bound1}
\ea
This inequality is in fact mathematically rigorous, since
$(\rm prefactor)\sum c_{\j',\Delta'}\leq 1$ due to convergence of the Euclidean OPE,
and $\frac{1-\sqrt{\rho}}{1+\sqrt{\rho}}\leq e^{-2\sqrt{\rho}}\leq e^{-\sqrt{z}}$.
We stress the importance of focusing on the double discontinuity,
otherwise double-trace operators (which exist below the gap) would also contribute.

Changing variables to $(z,\zbar)=\sigma e^{\pm t}$ with $\sigma$ small,
the heavy contribution to the inversion formula (\ref{fg}) becomes, using (\ref{G_gegen}),
\be
 c^t(\j,\Delta)\big|_{\rm heavy} = C
 \int_0^1 \frac{d\sigma}{\sigma}\,\sigma^{j-1}
\int_{-\infty}^\infty dt \big|\sinh(t)\big|^{d-2} \tilde{C}_{\Delta+1-d}(\cosh t) \,\dDisc \GG\big|_{\rm heavy}\,, \label{inv_heavy}
\ee
where $C=\frac{\sqrt{\pi}\Gamma(\Delta-1)\kappa_{\j+\Delta}}{\Gamma(\frac{d-1}{2})\Gamma(\Delta-\frac{d}{2})}$
is some constant.
For $\Delta=\frac{d}{2}+i\nu$ along the physical contour of integration,
the $t$ integration is nonsingular but the $\sigma$ integral is strongly suppressed away from its lower endpoint;
plugging in the upper bound  from eq.~(\ref{upper_bound1}), $\dDisc \GG\leq e^{-2\sqrt{\sigma}\cosh(t)\Delta_{\rm gap}}$,
one gets bounds of the type
\be
 \Big|c(\j,\tfrac{d}{2}+i\nu)_{\rm heavy}\Big| \leq \frac{\#}{(\Delta_{\rm gap}^2)^{\j-1}} \label{bound}
\ee
where $\#$ is a universal (theory-independent) constant (that may depend on $\nu$).

To be clear, eq.~(\ref{bound}) does \emph{not} assume that any large-$N$ factorization or
even distinction between single- and multi- traces exists \emph{above} the gap, only that a full, unitary, theory exists
in the UV limit $\sigma\to 0$ at finite $N$, and that the double traces \emph{below} the gap are numerically suppressed (for the double-discontinuity)
due to the parameter $1/N$.  Also no statement is needed about the UV behavior of the correlator order by order in $1/N$.

To put this bound into perspective we can look at the case $\j=2$, where $c(2,\Delta)$
must contain the stress tensor pole.  We recall that the inversion integral (\ref{inv_heavy}) is justified for $\j>1$,
which includes this case.
The contribution to $c(2,\Delta)$ of a finite number of light $t$-channel operators,
as studied in the preceding section, only produce poles at double-trace twists, which are only near the stress tensor pole
if $\Delta_1+\Delta_2-(d-2)\approx 0$, that is if the external scalar operators nearly saturate the unitary bound, which
we do not expect to happen in strongly coupled theories.  Therefore the stress tensor residue ${\sim} 1/c_T\sim 1/N$
must be saturated by heavy $t$-channel operators (a similar conclusion is obtained
on the gravity side \cite{Polchinski:2002jw}), so (\ref{bound}) gives a bound of the type
$c_T\geq \#\Delta_{\rm gap}^2$ with a calculable coefficient.
From the gravity perspective, this is the statement (expected from unitarity) that new states must appear below the Planck scale.
(Similar parametric bounds were obtained in Mellin space \cite{Alday:2016htq}; the improvement here
stems from $\dDisc\GG$ being locally bounded, in contrast to the Mellin amplitude, which makes the argument nonperturbative in $1/N$.)

Since the double discontinuity is positive definite and locally bounded,
one can use the stress tensor contribution to control its overall normalization,
thus rewriting eq.~(\ref{bound}) as
\be
 \Big|c(\j,\tfrac{d}{2}+i\nu)_{\rm heavy}\Big| \leq \frac{1}{c_T}\frac{\#}{(\Delta_{\rm gap}^2)^{\j-2}}\,, \label{bound1}
\ee
again with some computable coefficient.  Again this is valid for $\j>1$.

\begin{figure}
\be\begin{array}{ccc}
\hspace{0mm}\def\svgwidth{30mm}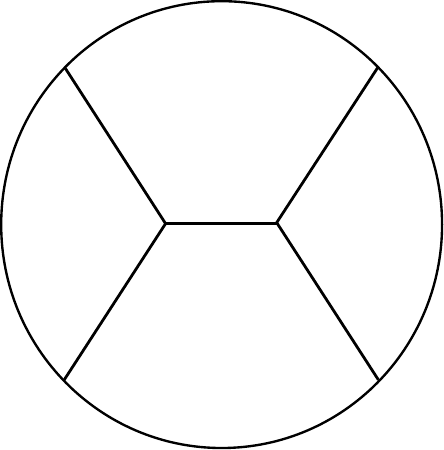
& \hspace{3mm}
\raisebox{14mm}{\resizebox{5mm}{!}{$\Rightarrow$}}
\hspace{3mm}\def\svgwidth{30mm}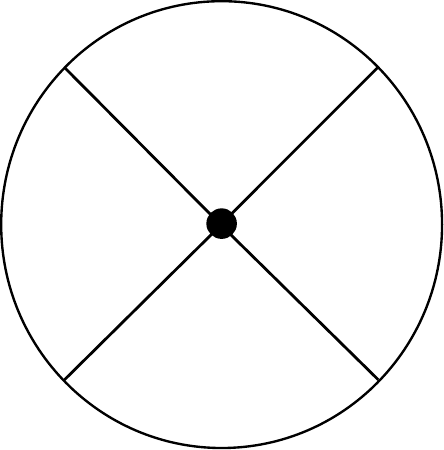
&
\hspace{25mm}\def\svgwidth{41mm}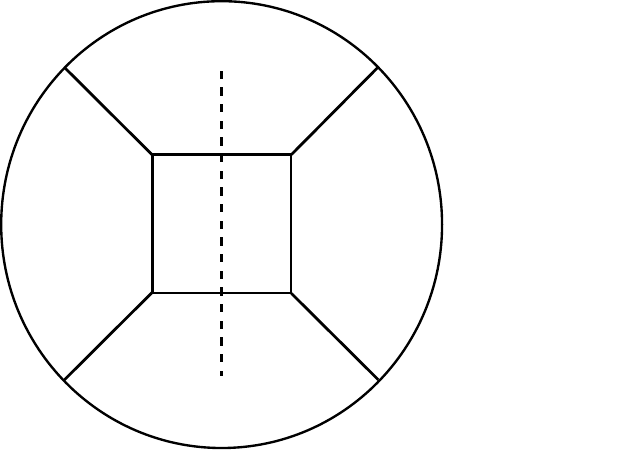
\\ \\ &\hspace{-31mm}\mbox{(a)}&\hspace{14mm}\mbox{(b)}
\end{array}\nonumber\ee
\caption{(a) Integrating out a heavy field in AdS space produces local higher-derivative
interactions, suppressed by powers of the heavy mass.
 (b) The double-discontinuity of a one-loop correction is equal to a product of trees.}
\label{fig:ads}
\end{figure}

This bound can be compared to expectations from effective field theory in AdS.
In this setup, ``heavy'' operators represent fields of large mass in AdS units, which can be integrated
out when computing correlators of light fields, as depicted in fig.~\ref{fig:ads}.
This produces a series of higher-derivative corrections suppressed by inverse power of the heavy mass,
of the schematic form $(\partial^2/\Delta_{\rm gap})^m \phi^4$ with various contractions.
As explained originally in \cite{Heemskerk:2009pn}, the Witten diagrams associated with these
interactions give rise to solutions to the crossing equation which are supported by double trace primaries with
finitely many spins, $\j=0,\ldots m$. 

We see that the bound (\ref{bound1}) coincides with the expected optimal one
in the case where derivatives are organized to produce the maximal angular momentum (in the channel under consideration):
each extra factor of $\partial^2$ is then suppressed by $\Delta_{\rm gap}^2$.
In general, the bound (\ref{bound1}) is however weaker because it detects the angular momentum of an interaction
rather than its mass dimension.

This is still highly constraining due to crossing symmetry, because it is not possible
for a four-scalar interaction to have very many derivatives without having large spin in at least one of the $s$-, $t$- or $u$-channel.
Therefore, except for finitely many exceptions (such as the six-derivative interaction represented by the flat space amplitude $stu$,
which has spin 2 in all channels), the Regge limit constraint (\ref{bound1}) proves that the coefficients of all higher-derivative interactions contributing to a four-point correlator must be small and decay with increasing
dimension, as conjectured in \cite{Heemskerk:2009pn}.  We also expect the bounds to be more constraining for external operators in spin,
due to the restrictions on their local self-interactions.

We thus believe that the bound (\ref{bound1}) goes a long way toward establishing that all large-$N$ theories with a large gap
admit a local gravity dual, although it will be important to improve it toward the expected optimal bounds, which
will presumably require information from limits other then the Regge limit.  Also it will be important to gain better control over the low spins -- for example the spin 0 interaction $\phi^4$ -- which are generally expected to have also small coefficients in AdS/CFT
but over which we provide no control here.  Another important question is whether a theory dual to a CFT
with a large gap can have a light spin-two particle beyond the graviton, which is not expected on the gravity side but not ruled out by the present arguments.  (Of course, with supersymmetry, there can be more restrictions; for instance the constraint implemented in \cite{Rastelli:2016nze},
that the correlator cannot grow faster in the Regge limit than spin-two exchange, fully determined the correlators
in the $\mathcal{N}=4$ theory and is rigorously justified by the above bounds.)

Finally, let us briefly comment on loop corrections at large $N$.
Since the double discontinuity extracts, roughly speaking, the coefficient of $\log^2(1-\zbar)$,
double-trace operators start to contribute to it at the one-loop order, but \emph{only in a way proportional to the leading
$1/N$ anomalous dimension of $t$-channel operators}, which can be extracted from tree-level amplitudes.
(A similar conclusion was reached recently in Mellin space \cite{Aharony:2016dwx}.) Note that, because
of mixing among double traces, one has to sum over all the intermediate primaries $\OO$, $\tilde\OO$.
All we would like to add here is that, since the amplitude $\mathcal{M}$ already extracts one discontinuity (see eq.~(\ref{defM})),
the double-discontinuity at one-loop can be written in a very suggestive form: ${\rm Im}\,\mathcal{M} =\mathcal{M}\otimes \mathcal{M}$ (the tensor product sign representing multiplication in $(\j',\Delta')$ space, divided by free theory OPE coefficient of the $t$-channel intermediate operators, $f^2_{\OO\tilde\OO[\OO\tilde\OO]}$).

\pagebreak
\section{Conclusion}\label{sec:conclusion}

The aim of this paper was to present a mathematical formula. The formula is in eq.~(\ref{fg}).

Given a four-point correlator in a conformal field theory, the formula returns the operator dimensions and
OPE coefficients which lead to it. This data is an analytic function of the spin of the exchanged operator,
and the formula, similarly to a dispersive representation, quantifies the consequences of this fact.
The input is that the Lorentzian correlator admits a sensible high-energy Regge limit,
which physically is a consequence of crossing symmetry and of the positivity of Euclidean OPE data.
A simplified formula, which exploits only the SL${}_2$(R) conformal symmetries of a null line,
is given in eq.~(\ref{C_coll}).

We have illustrated the formula in a number of applications and tests.  In section \ref{sec:spin}
we showed how it concisely encodes a body of existing results on  operators with large spin.
Most importantly, it explains conceptually why these operators organize into analytic families in the first place.
It provides convergent sums, instead of asymptotic series in $1/\j$. We hope that this will be of great help
to control its errors.

In theories with a large-$N$ factorization, the formula is saturated by single-trace operators.
In the case where the spectrum is sparse this enables to bound OPE coefficients
of double-trace operators of spin larger than 2, see eq.~(\ref{bound1}).
These bounds match the expectations from a dual AdS theory that would be local down to distances of order $1/\Delta_{\rm gap}$ in AdS units.
We believe that this goes a long way toward proving that any large-$N$ CFT with a large gap has a local gravity dual,
as originally conjectured in \cite{Heemskerk:2009pn}.
They do not saturate the expectations yet however, as
was expected since they use only the Regge limit and not the more general high-energy fixed-angle limit.

We feel that a lot remains to understand, and that the tool could be useful for other questions.
Analyticity in spin is only guaranteed to apply to OPE data for spin $\j\geq 2$, and
it would be important to better understand spins 0,1 (which in this paper are only covered by the Euclidean inversion formula
(\ref{eucl_inversion})). One might ask for instance if strong scalar self-interactions are possible in theories with AdS gravity dual. 
Also it would be nice to better understand why our main formula eq.~(\ref{fg}) exists at all,
due to the overconstrained nature of the system we solved.
This could also help generalize the formula to external operators with spin; such a generalization
could shed new light on the relation between the $a$ and $c$ central charges in theories with AdS gravity dual
\cite{Camanho:2014apa,Afkhami-Jeddi:2016ntf}.
In two dimensions, extending the formula to Virasoro blocks would also be interesting.
Our finding that correlators are analytic in spin for $\j\geq 2$ squares well with the recent conjecture of 
\cite{Collier:2017shs}, that the only theories with Virasoro primaries of bounded spin have maximal spin 0
(and are Liouville theory specifically),
although from the present perspective it is not immediately clear why spin 1 should not be possible.
In the context of the $1/N$ expansion, it could be fruitful to pursue the analogy with $S$-matrix unitarity
and higher-point correlators sketched at the end of the preceding section
-- this could help organize multi-twist operators in more general theories.  In general theories, it would be exciting to implement
the error control on the large-spin expansion suggested in section \ref{sec:spin},
so as to be enable using these analytic predictions within the numerical bootstrap.

Finally, it is worth remembering that the Froissart-Gribov formula originated over 50 years
ago in the context of the $S$-matrix bootstrap.  Have all its applications been found?

\acknowledgments
I would like to thank David Simmons-Duffin and the participants of the Simons Collaboration on the Non-perturbative Bootstrap
meetings held in Lausanne, January 2017, and Princeton, March 2017,
for comments and stimulating discussions.

\begin{appendix}
\section{Conformal blocks in general dimensions}\label{app:blocks}

Conformal blocks are characterized by the dimensions and spin of the exchanged primary operator.
This data is encoded in the quadratic and quartic Casimir invariants
of the conformal algebra \cite{Dolan:2011dv,Hogervorst:2013kva}:
\be\begin{aligned}
 c_2&= \tfrac12\left[\j(\j+d-2)+\Delta(\Delta-d)\right],\\
 c_4&= \j(\j+d-2)(\Delta-1)(\Delta-d+1)\,, \label{Casimirs}
\end{aligned}\ee
which are the eigenvalues of the following differential operators:
\be\begin{aligned}
 C_2&= D_z+D_{\zbar}+(d-2)\frac{z\zbar}{z-\zbar}\left[(1-z)\partial_z-(1-\zbar)\partial_{\zbar}\right],\\
 C_4&= \left(\frac{z\zbar}{z-\zbar}\right)^{d-2} (D_z-D_{\zbar})\left(\frac{z\zbar}{z-\zbar}\right)^{2-d}(D_z-D_{\zbar})\,.
\label{Casimir_ops}\end{aligned}\ee
Here
\be
D_z=z^2\partial_z(1-z)\partial_z-(a+b)z^2\partial_z-abz
\ee
and $a=\tfrac12(\Delta_2-\Delta_1)$, $b=\tfrac12(\Delta_{3}-\Delta_4)$ as in the main text.

In even spacetime dimensions, explicit closed-form solutions can be obtained in terms of hypergeometric functions,
as given in eqs.~(\ref{blocks24}).  In the general case, it is necessary to rely on other methods such
as the various series expansions discussed in this appendix.  

Since we will be interested in analytic continuations, it will be useful to consider the most general solution
to these equations. These are most concisely described when $\j$ and $\Delta$
are generic (such that $\j$ and $\j\pm \Delta$ are non-integer) -- non-generic cases can then be obtained as limits.
One can then choose solutions that are pure power laws in the limit $0\ll z\ll \zbar\ll 1$, labelled by their exponents:
\be
 \gpure_{\j,\Delta}(z,\zbar) = z^{\frac{\Delta-\j}{2}}\zbar^{\frac{\Delta+\j}{2}}\times (1 + \mbox{integer powers of $z/\zbar,\zbar$})\,. \label{gpure}
\ee
There are in fact eight independent solutions, which are obtained from the above
by acting with the three independent $Z_2$ symmetries of the Casimir eigenvalues (\ref{Casimirs}):
\be
 \j{\longleftrightarrow}2-d-\j; \qquad \Delta\longleftrightarrow d-\Delta;\qquad \Delta \longleftrightarrow 1-\j\,. \label{symmetries}
\ee

\subsection{Expansions around the origin $z,\zbar\to 0$}

In the limit that $z,\bar{z}\to 0$, the dependence on their ratio 
is controlled by the Gegenbauer differential equation with $x=\cos\theta=\frac12(\sqrt{z/\zbar}+\sqrt{\zbar/z})$:
\be
 \gpure_{\j,\Delta} = (z\zbar)^{\frac{\Delta}{2}}(f_\j(x) + O(z\zbar)),\qquad
\left[ (1-x^2)\partial_x^2 -(d-1)x\partial_x+\j(\j+d-2)\right]f_\j(x)=0\,.
\ee
This is the $d$-dimension generalization of spherical harmonics, e.g. Legendre polynomials.
This is physically expected since in Euclidean kinematics $x$ is the cosine of an angle.
The pure power solutions corresponding to (\ref{gpure})
can be written as
\be
 f_{\j}(x) = (2x)^{\j} {}_2F_1\big(\tfrac{-\j}{2},\tfrac{1-\j}{2},2-\j-\tfrac{d}2,\tfrac{1}{x^2}\big)\,. \label{fj_hyper}
\ee
On the other hand, there is always a solution analytic around $x=1$, known as Gegenbauer polynomial
(normalized here to $\tilde{C}_j(1)=1$)
\be
\tilde{C}_\j(x)\equiv \frac{\Gamma(\j+1)\Gamma(d-2)}{\Gamma(\j+d-2)}\,C^{d/2-1}_\j(x) =
{}_2F_1\big(-\j,\j+d-2,\tfrac{d-1}{2},\tfrac{1-x}{2}\big)\,.
 \label{gegen}
\ee
It is, of course, only a polynomial when $\j$ is an integer.
Comparing its large-$x$ asymptotics with the normalization (\ref{normalization}),
we conclude that the blocks behave near the origin like
\be
 G_{\j,\Delta}(z,\zbar) = \frac{2^{3-d}\sqrt{\pi}\Gamma(\j+d-2)}{\Gamma\big(\tfrac{d-1}{2}\big)\Gamma\big(\j+\tfrac{d-2}{2}\big)}
 (z\zbar)^{\frac{\Delta}{2}}\left( \tilde{C}_\j(x) + O(z\zbar)\right)\,. \label{G_gegen}
\ee
More generally, the Gegenbauer polynomials can be written as a sum of two pure power solutions.
This gives an exact decomposition of the regular solution $G$ used in the main text:
\be
 G_{\j,\Delta} (z,\zbar)= \gpure_{\j,\Delta}(z,\zbar) +
\frac{\Gamma(\j+d-2)\Gamma\big(-\j-\tfrac{d-2}{2}\big)}{\Gamma\big(\j+\tfrac{d-2}{2}\big)\Gamma(-\j)}
\gpure_{-\j-d+2,\Delta}(z,\zbar). \label{G_from_gpure}
\ee
This will be useful since the pure power solutions $\gpure_{\j,\Delta}$ have simpler asymptotic expansions
and analytic continuations.  (This formula is valid for generic dimension. The limit to even spacetime dimension can be singular
for integer spins, but this does not appear to affect our final formulas.)

\subsubsection*{Orthonormality of conformal blocks}\label{app:ortho}

Interchanging the order of integrations in the Euclidean inversion formula (\ref{eucl_inversion}),
with $\Delta=\frac{d}{2}+i\nu$, one finds the following integral:
\be
 I_{\j,\nu; \j',\nu'} \equiv \int d^2z \,\mu(z,\zbar)\,g(z,\zbar)\, F_{\j,\frac{d}{2}+i\nu}(z,\zbar)F_{\j',\frac{d}{2}+i\nu'}(z,\zbar)\,.
\ee
It is easy to verify that, at least when $a,b$ are small enough,
the functions $F$ are sufficiently regular near $0$, $1$, $\infty$ that the Casimir operators are self-adjoint.
For example, when $a=b=0$, the $F$'s have at most logarithmic singularities near $z=1$, and the double-derivatives in $D_z$
come with a factor $(1-z)\partial_z^2$ which ensures that all boundary terms near $z=1$ arising from
integration-by-parts vanish.
When $a,b\neq 0$, the integral may become divergent near $z=1$, but as discussed in section \ref{ssec:eucl}
any contribution from the region $z,\zbar\approx 1$ has a specific $\j,\nu$
dependence which removes any possible physical consequence.

Self-adjointness of the quadratic and quartic Casimirs implies that the above
integral vanishes unless both blocks have the same eigenvalues.
For real $\nu$ this forces $\j=\j'$ and $\nu=\pm \nu'$. The latter is a distributional term
which can only appear if the integral develops a singularity,
which in turn can only come from near the origin where the behavior of the functions is
similar to that of a Mellin transform:
\be
\int_0 \frac{dr}{r} r^{i(\nu\pm \nu')} = \pi\delta(\nu\pm \nu')+ \mbox{non-singular}
\ee
The angular integral, thanks to the appropriate factor for spherical harmonics in $d$ dimensions
which arises from the measure (\ref{measure}), $|(z-\zbar)|^{d-2}\propto (\sin\theta/2)^{d-2}$,
takes the form of the of the orthogonality relation of Gegenbauer polynomials and forces the spins to be equal:
\begin{flalign}
&&\int_{-1}^1 dx (1-x^2)^{\frac{d-3}{2}} \tilde{C}_\j(x)\tilde{C}_{\j'}(x) &=
 \delta_{\j \j'} \frac{2^{d-2} \Gamma(\j+1)\Gamma(\tfrac{d-1}{2})^2}{\Gamma(\j+d-2)(2\j+d-2)}& (\j,\j' =\mbox{integers})\,.
\end{flalign}
By combining this with the normalization of the single-valued functions $F$ in eq.~(\ref{F_single}),
we obtained the normalization factor in the Euclidean inversion formula (\ref{eucl_inversion}):
\be
\mathcal{N}(\j,\Delta)=\frac{\Gamma\big(\j+\tfrac{d-2}{2}\big)\Gamma\big(\j+\tfrac{d}{2}\big)K_{\j,\Delta}}{
2\pi\,\Gamma(\j+1)\Gamma(\j+d-2)K_{\j,d-\Delta}}\,. \label{eucl_normalization}
\ee

\subsubsection*{Subleading terms and poles of conformal blocks}

Subleading terms in $z\zbar$ can be worked out systematically using the methods of \cite{Hogervorst:2013sma}:
one postulates an expansion in terms of powers times Gegenbauer polynomials,
\be
 \gpure_{\j,\Delta}(z,\zbar) = \sum_{m=0}^\infty (z\zbar)^{\frac{\Delta+m}{2}}
 \sum_{k=-m}^m A_{k,m}
  \tilde{C}_{\j+k}(x)\,. \label{origin_expansion}
\ee
The quadratic Casimir equation
then gives a first order recurrence relation
for the coefficients $a_{k,m}$;
the spin changes only by one unit at each step
due to the addition properties of angular momentum.
We reproduce this recursion relation here in the more general
case where $a,b\neq 0$. It is seeded by $A_{k,0}=\delta_{k,0}$ times the prefactor in eq.~(\ref{G_gegen})
and reads:
\be
2\big(c_2(\j+k,\Delta+m)-c_2(\j,\Delta)\big) A_{k,m} =
\gamma^+_{\Delta+m-1,\j+k-1}A_{\j+k-1,m-1}+\gamma^-_{\Delta+m-1,\j+k}A_{\j+k+1,m-1}
\ee
where
\be
 \gamma^+_{E,\j} = \frac{(\j{+}d{-}2)(E{+}\j{+}2a)(E{+}\j{+}2b)}{2\j{+}d{-}2},\qquad
 \gamma^-_{E,\j} = \frac{\j(E{-}\j{-}d{+}2{+}2a)(E{-}j{-}d{+}2{+}2b)}{2\j{+}d{-}2}\,.
\ee
This recursion relation is valid whether or not $\j$ is integer
and holds equally for $G_{\j,\Delta}$ and $\gpure_{\j,\Delta}$,
one simply has to replace
$\tilde{C}_\j(x)$ by $\frac{\Gamma(\j+\frac{d-2}{2})}{\Gamma(\j+d-2)}f_\j(x)$ in the latter case.

Using this recursion relation, it is possible to determine the poles in the conformal blocks $G_{\j,\Delta}$
as a function of $\Delta$. 
Since the denominators come from the Casimir, it is easy to check that the only poles are at $\Delta=\j+d-2-m$,
where $m=0,1,2,\ldots$. This has a simple physical interpretation since $\Delta\geq\j+d-2$ is the unitarity bound
(for generic spin $\j$): the poles appear when the unitarity bound is crossed.
The residue must be one of the solutions related by the symmetries (\ref{symmetries}).
Working out the proportionality factor we find that the following combination is pole-free for $\Delta>d/2$:
\be
 G_{\j,\Delta}(z,\zbar) - r_{\Delta+1-d,\j+d-1} G_{\Delta+1-d,\j+d-1}(z,\zbar),
\ee
where $r_{\j,\Delta}$ is a messy-looking product of $\Gamma$-functions (with
$x=\Delta-j-d+2$):
\be
\label{r}
r_{\j,\Delta} = \frac
 {\Gamma(\Delta-1)\Gamma(\Delta+2-d)}{\Gamma\big(\Delta-\tfrac{d}{2}\big)\Gamma\big(\Delta-\tfrac{d-2}{2}\big)}
 \frac{\Gamma\big(\j+\tfrac{d-2}{2}\big)\Gamma\big(\j+\tfrac{d}{2}\big)}{\Gamma(\j+1)\Gamma(\j+d-2)}
 \frac{\Gamma(2-x)\Gamma\big(a+\tfrac{x}{2}\big)\Gamma\big(b+\tfrac{x}{2}\big)}
 {\Gamma(x)\Gamma\big(a+\tfrac{2-x}{2}\big)\Gamma\big(b+\tfrac{2-x}{2}\big)}\,.
\ee

\subsection{Expansions around $z\to 0$ and monodromy under analytic continuation}
\label{app:cont}

Many applications involve the collinear limit $z\to 0$. 
The $\zbar$ dependence is then controlled by the conformal symmetry SL${}_2$(R) of a one-dimensional null line;
in particular the quadratic Casimir (\ref{Casimir_ops}) reduces to a hypergeometric equation whose solution
appeared already in the two- and four-dimensional blocks in eq.~(\ref{blocks24}):
\be
 \gpure_{\j,\Delta}(z,\zbar)\xrightarrow{z\to 0} z^{\frac{\Delta-\j}{2}} k_{\Delta+\j}(\zbar), \qquad k_{\beta}(\zbar) =
 \zbar^{\beta/2}\,{}_2F_1(\beta/2+a,\beta/2+b,\beta,\zbar).
\ee
In the limit $\zbar\to 1$, the hypergeometric function admits the standard expansion
\be
 \lim_{\zbar\to 1}k_\beta(\zbar) =
 \frac{\Gamma(a+b)\Gamma(\beta)}{\Gamma(\beta/2+a)\Gamma(\beta/2+b)}(1-\zbar)^{-a-b}
+ \frac{\Gamma(-a-b)\Gamma(\beta)}{\Gamma(\beta/2-a)\Gamma(\beta/2-b)}+\ldots\label{hypers_at_one}
\ee
where the dots stand for infinite towers of integer power corrections to the two terms.
This formula will be used to seed the double null limit $z\to 0,\zbar\to 1$ in the next subsection.

The second solution to the quadratic equation involves $k_{2-\beta}(\zbar)$
and controls the collinear limit of $\gpure_{1-\Delta,1-\j}(z,\zbar)$.
When we analytically continue, in particular going to the Regge sheet by taking $\zbar$ counter-clockwise around 1
(while retaining $z$ small), these two solutions mix.
The continuation, which can be worked out from (\ref{hypers_at_one}), reads
\be\begin{aligned}
 \gpure_{\j,\Delta}(z,\zbar)^{\circlearrowleft} &=  \gpure_{\j,\Delta}(z,\zbar)\left[1 -2i\frac{e^{-i\pi(a+b)}}{\sin(\pi (\j+\Delta))}\sin(\pi(\tfrac{\j+\Delta}{2}+a))\sin(\pi(\tfrac{\j{+}\Delta}{2}+b))\right]
\\ & \quad- \frac{i}{\pi} \,\gpure_{1-\Delta,1-\j}(z,\zbar)\frac{e^{-i\pi(a+b)}}{\kappa_{\j{+}\Delta}} \label{cont_regge}
\end{aligned}\ee
with $\kappa_\beta$ defined in eq.~(\ref{kappa}).
In the text we also need the continuation as $z$ goes counter-clockwise around $1$, with $\zbar$ fixed.
The trick is to do this in multiple steps, first interchanging $z$ and $\zbar$ in the pure power solutions.
By analyzing the hypergeometric function (\ref{fj_hyper}) near $x=1$,
we obtain the following connection formula, if $z$ is analytically continued
to the right of $\zbar$ in a counter-clockwise fashion (so that $x$ goes counter-clockwise around 1):
\be
 \gpure_{\j,\Delta}(z,\zbar) = \gpure_{-\j-d+2,\Delta}(\zbar,z) \frac{e^{-i\pi\frac{d-2}{2}}\Gamma(-\j-\tfrac{d-2}{2})\Gamma(1-\j-\tfrac{d-2}{2})}{\Gamma(-\j)\Gamma(3-\j-d)} + \gpure_{\j,\Delta}(\zbar,z)\frac{e^{i\pi \j}\sin(\pi\tfrac{d-2}{2})}{\sin\big(\pi(\j+\tfrac{d-2}{2})\big)}\,. \label{cont_spin}
\ee
The continuation of $\gpure_{j,\Delta}(z,\zbar)$ counter-clockwise around $z=1$ can then be obtained simply
by applying (\ref{cont_spin}) followed by (\ref{cont_regge}) and then (\ref{cont_spin}) again.
(We caution the reader that the limit of integer spin should be approached with care in the preceding formula,
since the large solution $f_\j(x)$ diverges in this limit and both terms contribute.)

For single-valued combinations, the two ways of reaching the Regge sheet by rotating either $z$ or $\zbar$ counter-clockwise around 1
should give the same result. The preceding two formulas can therefore be used to confirm single-valuedness of the combination
$F_{\j,\Delta}$ in eq.~(\ref{F_single}).

\subsubsection*{Subleading terms}

To expand in subleading powers of $z$ one can proceed following methods similar to \cite{Dolan:2003hv}.
In fact in the main text what we really need is the expansion of the block times measure:
\be
z^{1-d}\left(\frac{\zbar-z}{\zbar}\right)^{d-2}\,(1-z)^{a+b}\,\gpure_{\Delta+1-d,\j+d-1}(z,\zbar) =
\sum_{m=0}^\infty z^{\frac{\j-\Delta}{2}+m} \sum_{k=-m}^m B^{(m,k)}_{\j,\Delta} k_{\Delta+\j+2k}\,(\zbar). \label{twist_expansion}
\ee
The coefficients $B^{(m,k)}_{\j,\Delta}$ (which are those entering eq.~(\ref{C_series}))
can then be obtained recursively using the quadratic Casimir equation.
To simplify the expansion it turns out to be convenient to consider a slightly different prefactor:
\be
z^{1-d}\left(\frac{\zbar-z}{\zbar}\right)^{\frac{d-2}{2}}\,\gpure_{\j,\Delta}(z,\zbar) =
\sum_{m=0}^\infty z^{\frac{j-\Delta}{2}+m} \,h^{(m)}_{\j,\Delta}(\zbar). \label{twist_expansion1}
\ee
We expand each term as a sum of SL${}_2$ blocks, $h_{\j,\Delta}^{(m)}(\zbar)=\sum_{k=-m}^m h_{\j,\Delta}^{(m,k)} k_{\beta+2k}(\zbar)$
with $\beta=\Delta+\j$ and $\tau=\Delta-\j$, and find that the quadratic Casimir equation becomes
\be\begin{aligned}
&\sum_k \big(k(k+\beta-1)+m(m+\tau+1-d)\big)h_{\j,\Delta}^{(m,k)} k_{\beta+2k}(\zbar)
\\ &\qquad =
\left(\tfrac12\big(\tau-d\big)+m+a\right)\left(\tfrac12\big(\tau-d\big)+m+b\right) h_{\j,\Delta}^{(m-1)}(\zbar)
\\ & \qquad \quad + \tfrac14(d-2)(d-4)
\sum_{m'=1}^m\left( \frac{2m'}{\zbar^{m'}}-\frac{2m'-1}{\zbar^{m'-1}}\right) h_{\j,\Delta}^{(m-m')}(\zbar)\,.
\end{aligned}\ee
The right-hand side can be expressed as a sum over SL${}_2$ blocks using the recursion relation:
\be
\frac{1}{\zbar} k_\beta(\zbar) = k_{\beta-2}(\zbar)+ \left(\frac12-\frac{2ab}{\beta(\beta-2)}\right)k_\beta(\zbar)+ \frac{(a^2-\tfrac14\beta^2)(b^2-\tfrac14\beta^2)}{\beta^2(\beta^2-1)}k_{\beta+2}(\zbar)\,.
\ee
In this way the coefficients $h_{\j,\Delta}^{(m,k)}$ can be obtained recursively.  Multiplying by $(1-z/\zbar)^{\frac{d-2}{2}}(1-z)^{a+b}$
and exchanging $\j\leftrightarrow 1-\Delta$ to go from (\ref{twist_expansion1}) to (\ref{twist_expansion}) and re-expanding in $z$,
then gives the desired $B^{(m,k)}_{\j,\Delta}$ coefficients.
As simple examples we find $B^{(0,0)}_{\j,\Delta}=1$ and $B^{(1,1)}_{\j,\Delta}=-\frac{(d-2)(\j+2)}{2\j+d}$.

\subsection{Expansions around $\zbar\to 1$}\label{app:cross}

Operators with low twist and high spin in the $s$-channel are controlled by the behavior
of $t$-channel blocks $G(1-\zbar,1-z)$ in the limit $z\to 0,\zbar\to 1$. The argument of the block itself thus appraochs
$(0,1)$.
An expansion in powers of $z$ around this limit, that is, in powers of the second argument of the block,
was defined in (\ref{cross_expansion_text}) and corresponds to the twist expansion in the $s$-channel.
Reverting to $(z,\zbar)$ arguments it can be written equivalently as
\be
G_{\j,\Delta}(z,\zbar) = \sum_{i=1,2}\sum_{m=0}^\infty (1-\zbar)^{p_i+m}H_{\j,\Delta}^{(i),m}(z)\,.
 \label{cross_expansion}
\ee
The quadratic Casimir equation is singular in this limit $\zbar\to 1$, since the operator $D_{\zbar}$ reduces $m$ by one unit.
One concludes that there can be only two towers of terms, beginning with the two exponents
corresponding to the zero-modes of $D_{\zbar}$: $p_1=0$ and $p_2=-a-b$, which represent double-twist $s$-channel operators.
(When these blocks are used in $t$-channel, which is related by $1\leftrightarrow3$ to the $s$-channel, one should use that
$a|_{t-{\rm channel}}=\frac{\Delta_2-\Delta_3}{2}$ and $b|_{t-{\rm channel}}=\frac{\Delta_1-\Delta_4}{2}$. The $H_{\j,\Delta}^{(i),m}$
functions defined by eq.~(\ref{cross_expansion}) then become precisely those entering eq.~(\ref{cross_expansion_text}).)

The solutions are more conveniently described by decomposing the blocks into pure power solutions (\ref{G_from_gpure}).
Normalizing the contributions so that they have simple behavior at $z\to 0$,
this gives
\be
 \gpure_{\j,\Delta}(z,\zbar) = \sum_{i=1,2}\sum_{m=0}^\infty (1-\zbar)^{p_i+m}c^{(i)}_{\j+\Delta}\tilde{H}_{\j,\Delta}^{(i),m}(z)
 \label{cross_expansion1}
\ee
where in the $z\to 0$ limit, $\tilde{H}_{\j,\Delta}^{(i),m}(z)= z^{\frac{\Delta-\j}{2}}$ times a tower of integer powers,
and the coefficients, coming from the collinear expansion (\ref{hypers_at_one}), are
\be
 c^{(1)}_\beta = \frac{\Gamma(\beta)\Gamma(-a-b)}{\Gamma(\beta/2-a)\Gamma(\beta/2-b)}\,,\qquad
 c^{(2)}_\beta = \frac{\Gamma(\beta)\Gamma(a+b)}{\Gamma(\beta/2+a)\Gamma(\beta/2+b)}\,.
\ee
In the case where $a+b=0$, which occurs for identical operators, logarithms of $\zbar$ appear
and this need to be rewritten, following eq.~(\ref{cross_expansion_log_text}), as
\be\label{cross_expansion_log}
 \gpure_{\j,\Delta}(z,\zbar) =
 \sum_{m=0}^\infty (1-\zbar)^{m}
c^{\slog}_{\j+\Delta} 
 \left(\big(\tfrac12\log(1-\zbar)+c^{{\rm fin}}_{\j+\Delta}\big)\tilde{H}_{\j,\Delta}^{\slog,m}(z)+\tilde{H}_{\j,\Delta}^{\sreg,m}(z)\big)\right),
\ee
where $\tilde{H}^{\slog/\sreg,0}_{\j,\Delta}/z^{\frac{\Delta-\j}{2}}$ respectively tend to 1 or 0 as $z\to 0$.
The coefficients are now
\be
c^{\slog}_{\beta}= -\frac{2\Gamma(\beta)}{\Gamma(\beta/2-a)\Gamma(\beta/2+a)}\,,
\qquad
c^{{\rm fin}}_{\beta}= \tfrac12\psi(\beta/2-a)+\tfrac12\psi(\beta/2+a)-\psi(1)\,,
\ee
where $\psi(x)=\Gamma'(x)/\Gamma(x)$ is the polygamma function.

When this expansion is used in the $t$-channel, the expansion in powers of $z$ is dual to $1/\j$ corrections in the $s$-channel.
The coefficients are determined by the Casimir equations, but this is actually subtle,
since in this limit both the quadratic and quartic Casimirs mix the $m=0$ and $m=1$ terms. To resolve this,
we start from the equation for $C_4$ and substitute in the equation for $C_2$ to remove the offending lowering operators
$D_{\zbar}$; after some algebra we find:
\be
 c_4 G_{\j,\Delta}(z,\zbar) =
 \left(\frac{z\zbar}{z-\zbar}\right)^{d-2}\big(2D_z-Y-c_2+2-d\big)
\left(\frac{z\zbar}{z-\zbar}\right)^{2-d}\big(2D_z+Y-c_2\big)G_{\j,\Delta}(z,\zbar)\,,
\ee
where $Y=(d-2)\frac{z\zbar}{z-\zbar}\left[(1-z)\partial_z-(1-\zbar)\partial_{\zbar}\right]$ is the operator entering (\ref{Casimir_ops}).
The limit $\zbar\to 0$ is now regular and gives a closed equation for $m=0$:
\be\begin{aligned}
 c_4 \tilde{H}_{\j,\Delta}^{(i),0}(z)&=
 \left(\frac{z}{1-z}\right)^{d-2} \big(2D_z -\tilde{Y}-c_2 +2-d\big) \left(\frac{z}{1-z}\right)^{2-d}
 \big(2D_z +\tilde{Y}-c_2\big)\tilde{H}_{\j,\Delta}^{(i),0}(z)\,, \label{quartic_cross_channel}
\end{aligned}\ee
where now $\tilde{Y}=-(d-2)z(\partial_z + \frac{p_i}{1-z})$.
That the equation is quartic was to be expected since there needs to be 8 solutions, and there are only two exponents $p$.
In the logarithmic case $a+b=0$, one simply replaces $p_i$ by $d/d\log(1-\zbar)$.

We don't know whether this quartic differential equation can be solved in terms of more elementary functions.
In any case, it can be solved numerically, or also
straightforwardly as a power series in $z$: the above differential equation
directly translates into a fourth-order recurrence relation.
For example, focusing on the case where $d=3$ and $a=b=0$ as used in the main text,
and expanding in the variable $y=\frac{z}{1-z}$, we find that
\be\begin{aligned}
\tilde{H}_{\j,\Delta}^{\slog,0}(y)/y^{\frac{\Delta-\j}{2}} &= 
1+ \frac{(\Delta-\j-1)(\Delta(1-2\j)+\j)}{2(2\Delta-1)(2\j-1)} y + \big((\Delta-\j)^2-1\big)\times
\\&\hspace{-10mm} \times\frac{\big(\Delta^3(4\j^2-8\j+3)-\Delta^2(4\j^3-12\j^2+13\j-6)+\Delta \j^2+\j^2(\j-2)\big)}
{8(2\Delta-1)(2\Delta+1)(\Delta-\j)(2\j-3)(2\j-1)}y^2 +\ldots
\\
\tilde{H}_{\j,\Delta}^{\sreg,0}(y)/y^{\frac{\Delta-\j}{2}} &=
-\frac{y}{2(2\Delta-1)(2\j-1)} \\
& \qquad -\frac{\Delta^2(3-2\j)+2\Delta(3-2\j+2\j^2)+3-5\j+\j^2}{4(2\Delta-1)(2\Delta+1)(2\j-3)(2\j-1)}y^2 + \ldots \label{example_cross_series}
\end{aligned}\ee
In the special case $\j=0$, the equation factorizes, and correspondingly we find that the series simplifies
and can be summed exactly:
\be
 \tilde{H}_{0,\Delta}^{(1),0}(z) ={}_2F_1\big(\tfrac{\Delta}{2}+a,\tfrac{\Delta}{2}+b,\Delta-\tfrac{d-2}{2},z\big),
 \qquad
  \tilde{H}_{0,\Delta}^{(2),0}(z) =(1-z)^{-a-b}{}_2F_1\big(\tfrac{\Delta}{2}-a,\tfrac{\Delta}{2}-b,\Delta-\tfrac{d-2}{2},z\big).
\label{exact_j0}
\ee
In the limit $a+b\to 0$, either of these solution converges to $\tilde{H}_{0,\Delta}^{\slog,0}(y)$
as quoted in the main text.

Finally, we note that once the quartic equation for the $z$-dependence of the leading $\zbar\to 1$ term is solved,
it does not need to be used anymore since the quadratic equation expresses the subleading $(1-\zbar)$ terms
in terms of $z$-derivatives of the leading solution.

\section{A worked example: the $2D$ Ising model}
\label{app:Ising}

In this section we test the Lorentzian inversion formula on explicit correlators
of the two-dimensional critical Ising model.  This model contains two scalar Virasoro primaries:
$\sigma$ and $\epsa$, which have mass dimension $\Delta=\frac18$ and $\Delta=1$
and are odd and even under a $Z_2$ symmetry, respectively.
Their four-point correlators, with the conventions from subsection \ref{ssec:blocks},
are \cite{Belavin:1984vu} (expressions from \cite{Ardonne:2010hj}):
\be\begin{aligned}
 \GG^A&\equiv \GG_{\sigma\sigma\sigma\sigma} = \left|\frac{1}{(1-\rho^2)^{1/4}}\right|^2+\left|\frac{\sqrt{\rho}}{(1-\rho^2)^{1/4}}\right|^2,
& \qquad
 \GG^B&\equiv \GG_{\sigma\sigma\epsa\epsa}= \left|\frac{1+\rho^2}{1-\rho^2}\right|^2, \\
 \GG^C&\equiv \GG_{\sigma\epsa\epsa\sigma}= \left|\frac{\rho^{1/16}(1+6\rho+\rho^2)}{2^{7/8}(1-\rho)^2(1+\rho)^{1/8}}\right|^2,
 &\qquad
 \GG^D&\equiv \GG_{\epsa\epsa\epsa\epsa}= \left|\frac{1+14\rho^2+\rho^4}{(1-\rho^2)^2}\right|^2\,. \label{Ising}
\end{aligned}\ee
Note that $\GG^B$ and $\GG^C$ represent different channels of the same correlator.

Let us consider $\GG^A$ in detail, leaving the others to the reader (the others are also interesting as they manifest the divergences at $\zbar\to 1$ discussed below eq.~(\ref{C_residues}).
To compute its double discontinuity, we need to treat $\rho,\rhobar$ as independent variable and take $\rho\to 1/\rho$, either above or below the axis (see eq.~(\ref{dDisc})), which gives
\be
 \dDisc \GG^A(\rho,\rhobar) = \frac{1-\tfrac{1}{\sqrt{2}}(\sqrt{\rho}+\sqrt{\rhobar})+\sqrt{\rho\rhobar}}{(1-\rho^2)^{1/4}(1-\rhobar^2)^{1/4}}\,,
\ee
which here is to be evaluated in the range $0<\rho,\rhobar<1$
(thus differing from the formulas in the main text by $\rhobar\to1/\rhobar$)
Note that it is positive, as required.
To get the OPE coefficients, according to our main inversion formula (\ref{fg}),
we just need to integrate this against the two-dimensional (global) conformal blocks given in eq.~(\ref{blocks24}).
Since both the global blocks and Ising correlator take on a factorized form, we get a sum of factorized integrals:
\def\II#1#2#3{{I^{#1}_{#2}(#3)}}
\be\begin{aligned}\label{cA}
 c^A_{\j,\Delta} &= \big(1+(-1)^\j\big)\frac{\kappa_{\j+\Delta}}{2}\Bigg[\II{0}{-1/4}{\j{+}\Delta}\,\II{0}{-1/4}{\j{+}2{-}\Delta}+
 \II{1/2}{-1/4}{\j{+}\Delta}\,\II{1/2}{-1/4}{\j{+}2{-}\Delta}
\\&\hspace{15mm} -\tfrac{1}{\sqrt{2}}\II{1/2}{-1/4}{\j{+}\Delta}\,\II{0}{-1/4}{\j{+}2{-}\Delta}
-\tfrac{1}{\sqrt{2}}\II{0}{-1/4}{\j{+}\Delta}\,\II{1/2}{-1/4}{\j{+}2{-}\Delta}\Bigg]\,,
\end{aligned}\ee
where, in terms of the $\rho$-variables, using the measure given in (\ref{measrho}), the basic integral is
\be
\II{p_0}{p_1}{\beta} \equiv \int_0^1 \frac{d\rho\,(1-\rho^2)}{4\rho^2} k_\beta(\rho)
\rho^{p_0}(1-\rho^2)^{p_1}\,.
\ee
This looks very difficult because $k$ is an hypergeometric function with argument $\rho$, but in fact in the case that $a=b=0$
it can be written in terms of a hypergeometric with argument $\rho^2$:
\be
k_\beta(z)\big|_{a=b=0} = (4\rho)^{\beta/2}\,{}_2F_1\big(\tfrac12,\tfrac{\beta}{2},\tfrac{\beta+1}{2},\rho^2\big)\equiv k_\beta(\rho).
\ee
Changing variable to $u=\rho^2$ the integral can then be computed as a generalized hypergeometric function,
\be
\II{p_0}{p_1}{\beta} =
\frac{2^{\beta-3}\Gamma(p_1+2)\Gamma\big(\tfrac{\beta+2p_0-2}{4}\big)}{\Gamma\big(\tfrac{\beta+2p_0+4p_1+6}{4}\big)}
{}_3F_2\big(\tfrac12,\tfrac{\beta}{2},\tfrac{\beta+2p_0-2}{4}; \tfrac{\beta+1}{2},\tfrac{\beta+2p_0+4p_1+6}{4}; 1\big)\,.
\ee
The usual, discrete OPE
coefficients are then obtained from these partial wave coefficients by taking residues, according to (\ref{C_residues}):
\be
 c_{\j,\Delta}^A =
 -{\rm Res}_{\Delta'=\Delta}\left\{\begin{array}{l@{\hspace{10mm}}r}
 \c^A(\j,\Delta') & \mbox{($\Delta$ generic)}\\
 \c^A(\j,\Delta') -\frac{\Gamma(\j+2-\Delta')\Gamma^2(\frac{\Delta'-\j}{2})}
 {\Gamma(\Delta'-\j)\Gamma^2(\frac{\j+2-\Delta'}{2})} \c^A(\Delta'{-}1,\j{+}1) &
 (\Delta{-}\j=3,5,\ldots)
\end{array}\right.
\ee
For the first few OPE coefficients (up to dimension 7), for example, this formula gives:\footnote{We caution the reader that direct use
of Mathematica's \texttt{Residue[]} command may not always detect all poles of the ${}_3F_2$ functions.
The numbers quoted here were obtained by evaluating the residues numerically to high accuracy,
for example by taking $\Delta'-\Delta=10^{-50}$, and rationalizing the result.
}
\be\begin{aligned}
&c^A_{0,1}=\frac12\times\frac12=\frac14,\qquad c^A_{2,2} =\frac{1}{64},\qquad c^A_{4,4} = \frac{9}{40960},\qquad c^A_{0,4}=\frac{1}{4096},\\
&c^A_{4,5}=\frac{1}{65536},\qquad c^A_{6,6}=\frac{25}{3 670 016}\qquad c^A_{2,6}=\frac{9}{2621440},\qquad c^A_{6,7}=\frac{1}{1310720},\quad \ldots
\end{aligned}\ee
The extra $\frac12$ in the first case is due to the dimension being equal to $d/2$, as explained below eq.~(\ref{simple_radial_integral}).
We have checked that, upon substituting into the OPE sum (\ref{OPE}), these numbers reproduce the series expansion of the Euclidean correlator $G^A$!
(Using computer algebra we have checked the match up to dimension 15.)
This confirms the extraction of OPE data (with respect to the rigid, not Virasoro, conformal symmetry)
from the Lorentzian inversion formula, as an analytic function of dimension and spin.

We have also verified that the analytic result (\ref{cA}) agrees numerically with the Euclidean inversion integral (\ref{eucl_inversion}):
\be
 c^A_{\j,\Delta} = \mathcal{N}(\j,\Delta)\int_{|\rho|<1} d^2\rho\,\mu(\rho,\rhobar) \,G^A(\rho,\rhobar)\,F_{\j,\Delta}(\rho,\rhobar)\,.
\ee

\end{appendix}

\bibliographystyle{JHEP}
\bibliography{refs}

\end{document}